\documentclass{aa}
\usepackage{graphicx,psfig}
\usepackage{colortbl}
\usepackage{txfonts,amssymb}
\usepackage{color}

 \newcommand{\diff}{{\rm d}}

 \newcommand{\lp}{ \left(}
 \newcommand{\rp}{ \right)}

 \def\lesssim{\mathrel{\hbox{\rlap{\hbox{\lower4pt\hbox{$\sim$}}}\hbox{$<$}}}}
 \def\gtrsim{\mathrel{\hbox{\rlap{\hbox{\lower4pt\hbox{$\sim$}}}\hbox{$>$}}}}

\makeatletter
\let\@internalcite\cite
\def\cite{\@ifstar{\citeyear}{\citefull}}
\def\citefull{\def\astroncite##1##2{##1 ##2}\@internalcite}
\def\citeyear{\def\astroncite##1##2{##2}\@internalcite}
\def\citeau{\def\astroncite##1##2{##1}\@internalcite}
\def\citen{\def\astroncite##1##2{##1 (##2)}\@internalcite}
\def\possesivcite{\def\astroncite##1##2{##1's (##2)}\@internalcite}
\def\@citex[#1]#2{\if@filesw\immediate\write\@auxout{\string\citation{#2}}\fi
  \def\@citea{}\@cite{\@for\@citeb:=#2\do
    {\@citea\def\@citea{; }\@ifundefined
       {b@\@citeb}{{\bf ?}\@warning
       {Citation `\@citeb' on page \thepage \space undefined}}%
{\csname b@\@citeb\endcsname}}}{#1}}
\def\@cite#1#2{#1\if@tempswa , #2\fi}
\def\@biblabel#1{}

\def\etal{{et al.~}}

\def\kms{\>{\rm km}.{\rm s}^{-1}}
\def\Msun{\>{\rm M_{\odot}}}
\def\Lsun{\>{\rm L_{\odot}}}
\def\msun{${\rm M_{\odot}}$}
\def\cratio{$^{12}{\rm C}/^{13}{\rm C}$}

\makeatother
\authorrunning{Palacios et al.}
\titlerunning{Rotational mixing in low-mass stars. II}
\begin{document}

 \title{Rotational mixing in low-mass stars}
\subtitle{II. Self-consistent models of Pop~II RGB stars.}

\author{Ana Palacios$^{1}$, Corinne Charbonnel$^{2,3}$, Suzanne Talon$^{4}$
and Lionel Siess$^{1}$}

\offprints{Ana Palacios ; palacios@astro.ulb.ac.be or ana.palacios@cea.fr} 

\institute{
 Institut d'Astronomie et d'Astrophysique, Universit\' e Libre de Bruxelles
 Campus de la Plaine, Boulevard du Triomphe, CP 226, B-1050 Bruxelles, Belgium
 \and 
 Observatoire de Gen\`eve, 51 Chemin des Maillettes, CH-1290 Sauverny, Switzerland
 \and 
 Laboratoire d'Astrophysique de Toulouse-Tarbes - Observatoire
 Midi-Pyr\'en\'ees, 14 av, E. Belin, F-31400 Toulouse, France
 \and
  D\'epartement de Physique, Universit\'e de Montr\'eal, Montr\'eal PQ H3C 3J7,
 Canada
}

\abstract{In this paper we study the effects of rotation in low-mass, low-metallicity 
RGB stars. We present the first evolutionary models 
taking into account self-consistently the latest prescriptions 
for the transport of angular momentum 
by meridional circulation and shear turbulence
in stellar interiors as well as the 
associated mixing processes for chemicals     
computed from the ZAMS to the upper RGB.
We discuss in details the uncertainties associated with the physical
description of the rotational mixing and study carefully their effects on the rotation profile,
diffusion coefficients, structural evolution, lifetimes and chemical
signatures at the stellar surface.  We focus in particular on the various
assumptions concerning the rotation law in the convective envelope, the
initial rotation velocity distribution, the presence of $\mu$-gradients and
the treatment of the horizontal and vertical turbulence. 

This exploration leads to two main conclusions : (1) After the completion of the first
dredge-up, the degree of differential rotation (and hence mixing) is
maximised in the case of a differentially rotating convective envelope
(i.e., $j_{CE}(r) = cst$), as anticipated in previous studies. (2) Even
with this assumption, and contrary to some previous claims, the present treatment for the evolution of the rotation profile
and associated meridional
circulation and shear turbulence does not 
lead to enough mixing of chemicals to
explain the abundance anomalies in low-metallicity field and globular
cluster RGB stars observed around the {\em bump} luminosity.
This study raises 
questions that need to be addressed in a near future.
These include for example
the interaction between rotation and convection and the trigger of additional hydrodynamical instabilities.
}

\maketitle
\section{Abundance anomalies in RGB stars}\label{sec:intro}

The standard theory of stellar evolution\footnote{By this we refer to the modelling 
of non-rotating, non-magnetic stars, in which 
convection and atomic diffusion are the only transport processes considered.} 
predicts that the surface chemical composition
of low-mass stars is modified on the way to the red giant branch (RGB) during 
the so-called first dredge-up (hereafter 1st DUP; Iben \cite{Iben65}).
There, the expanding stellar 
convective envelope (hereafter CE)
deepens in mass, leading to the dilution of the surface material within
regions that have undergone partial hydrogen burning on the earlier 
main sequence (hereafter MS). 
Qualitatively, this leads to the decrease of the surface abundances of the
fragile LiBeB elements and of $^{12}$C, while those of $^3$He, $^{13}$C and
$^{14}$N increase.
Abundances of O and heavier elements remain essentially unchanged. 
Quantitatively, these abundance variations depend on the stellar mass
and metallicity (e.g., Sweigart, Greggio \& Renzini, \cite{Sweigart89}; Charbonnel \cite{CC94};
Boothroyd \& Sackmann \cite{BS99}).
After the 1st DUP, the CE withdraws while the hydrogen burning
shell (hereafter HBS) moves outward in mass. Within the standard framework 
no more variations of the surface abundance pattern are expected until the star 
reaches the asymptotic giant branch.

Observations sampling the evolution from the turn-off to the base of the RGB 
in open clusters and in the galactic field stars have validated 
these predicted surface abundances variations up to the completion of
the 1st DUP\footnote{One has of course 
to take into account possible variations of the surface abundance of lithium
occurring in some cases already on the MS. This discussion is 
however out of the scope of this paper (see e.g. Charbonnel, Deliyannis \& 
Pinsonneault \cite{CDP00} and Palacios et al. \cite{PTCF03}).} 
(e.g. Gratton \etal \cite{Gratton00}).
However observational evidence have accumulated of a second and distinct
mixing episode which is not predicted by standard models and which occurs
in low-mass stars after the end of the 1st DUP, and more precisely at the
RGB {\em bump}.

The determination of the carbon isotopic ratio $^{12}$C/$^{13}$C (hereafter
{\sl CIR}) for RGB stars in open clusters with various turn-off masses
(Gilroy \cite{Gilroy89}) provided the first pertinent clue on this process.
It was indeed shown that bright RGB stars with initial masses lower than
$\sim 2-2.5~{\rm M}_{\odot}$ exhibit {\sl CIR} considerably lower than
predicted by standard models after the 1st DUP.  Thanks to data collected in stars sampling the
RGB of M67 (Gilroy \& Brown \cite{GB91}), it clearly appeared that
observations deviated from standard predictions just at the so-called RGB
{\em bump} (Charbonnel \cite{CC94}).  The Hipparcos parallaxes allowed to
precisely determine the evolutionary phase of large samples of field stars
with known {\sl CIR}. These stars were found to behave similarly as those
in M67, e.g. presented unpredicted low {\sl CIR} appearing at the RGB {\em
bump} luminosity (Charbonnel, Brown \& Wallerstein \cite{CBW98}; Gratton et
al. \cite{Gratton00}).

On the other hand the region around the {\em bump} has also been probed for two 
globular clusters (GCs). In NGC 6528 and M4 again, the {\sl CIR} drops 
below the 1st DUP standard predictions just at the RGB {\em bump} 
(\cite{Shetrone03a,Shetrone03b}).
Moreover, all the brightest RGB stars observed so far in globular clusters 
exhibit {\sl CIR} close to the equilibrium value of the CN cycle.

During this second mixing episode surface abundances of other chemical
elements are also affected both in field and GCs giants : Li decreases at
the RGB {\em bump} (Pilachowski, Sneden \& Booth \cite{Pila93}; Grundahl et
al. \cite{Grundahl02}). C decreases while N increases for RGB stars brighter than the
{\em bump} (Gratton et al. \cite{Gratton00}; Bellman et al. \cite{Bellman01} and references
therein), confirming the envelope pollution by CN processing.
In the case of GCs, the picture is however blurred by the probable non-negligible
dispersion of the initial [C/Fe].
As far as lithium, carbon isotopes and nitrogen are
concerned, the abundance variations on the upper RGB
have similar amplitudes in field and globular cluster giants 
(Smith \& Martell \cite{Smith03}).
The finding that the so-called super Li-rich giants (Wallerstein 
\& Sneden 1982) all lie either at the RGB {\it bump} or on the
early-AGB (Charbonnel \& Balachandran \cite{CB00}), certainly indicates the
occurrence of an extra-mixing episode at these evolutionary points. The
trigger of this mixing episode has been suggested to be of
external nature (\cite{DH04}), but is much likely
related to the aforementioned second mixing episode, which would start with
a Li-rich phase as proposed by Palacios et al. (\cite{PCF01}).

For more than a decade it has been known that in addition to the elements
discussed previously, O, Na, Mg and Al also show variations in GC red
giants (Kraft et al. \cite{Kraft93}; Ivans et al. \cite{Ivans99}; Ramirez
\& Cohen \cite{RC02}).  As in the case of lighter nuclei, an {\em in situ}
mixing mechanism was frequently invoked to explain these abundance
anomalies, and in particular the O-Na anti-correlation. For a long time
this specific pattern could only be observed in the brightest GC RGB
stars. However, O and Na abundances have recently been determined with
8-10m class telescopes in lower RGB and in turn-off stars for a couple of
GCs (for recent reviews see Sneden 2005 and Charbonnel 2005 and references
therein) revealing exactly the same O-Na anti-correlation as in bright
giants. This result is crucial. Indeed, the NeNa-cycle does not operate in
MS low-mass stars, as the involved reactions require high temperatures that
can only be reached on the RGB. The existence of the same O-Na
anti-correlation on the MS and on the RGB in these clusters thus proves
that this pattern does not result from
self-enrichment. The recent determination of oxygen isotopic ratios in a few RGB
stars with low {\sl CIR} (Balachandran \& Carr \cite{Bala03})
reinforces this result. These objects indeed present high
$^{16}$O/$^{17}$O and $^{16}$O/$^{18}$O ratios, in agreement with
extensive CN-processing but no dredge-up of ON-cycle material. The O-Na
anti-correlation is generally assumed to be of primordial origin, even though it
has been proved difficult to find stellar candidates able to produce it
(\cite{TDCC05},\cite{DH03}). Let us finally mention the peculiar case of M13
bright giants, where O, Na, Mg and Al abundances appear to vary with
luminosity. In this cluster the observed O-Na anti-correlation could thus be the
result of a superimposition of self-enrichment with a primordial pattern
(\cite{Johnson05}).

In short, observations provide definitive clues on an additional mixing episode
occurring in low-mass stars after the end of the 1st DUP.
This process appears to be universal and independent of the stellar environment :
it affects more than 95$\%$ of low-mass stars (Charbonnel \& do Nascimento \cite{CdN98}), 
whether they belong to the field, to open or globular clusters. 
Some indications of such a process have also been detected in the brightest RGB stars
of external galaxies like the LMC (Smith et al. \cite{Smith02}) and Sculptor (Geisler et al. \cite{G05}). 
Its signatures in terms of abundance anomalies are clear :
the Li and $^{12}$C abundances as well as the $^{12}$C/$^{13}$C ratio drop
while the $^{14}$N and the $^{16}$O/$^{18}$O ratio increase.
These data attest the presence of a non-standard mixing process connecting the stellar 
convective envelope with the external layers of the HBS
where CN-burning occurs. Last but not least, they indicate that 
the effects of this process on surface abundances appear when the star
reaches the RGB {\em bump}.

Why should the RGB {\em bump} be such a special evolutionary point in the
present context? 
After the completion of the 1st DUP, the CE retreats leaving a
discontinuity of mean molecular weight (or $\mu$-barrier) at the mass coordinate of its maximum
penetration. 
Subsequently, when the HBS eventually crosses this
discontinuity, the star suffers a structural re-adjustment due 
to the composition changes (more H is made available in the burning
shell). The resulting alteration of the energetics 
causes a momentary decrease of the stellar luminosity. 
This results in a higher probability for finding a star in this brightness bin,
and translates into a {\em bump} in the luminosity functions of globular clusters.
Standard theory and observations nicely agree on the size and the location
of the {\em bump} in the HRD (e.g., Zoccali et al. \cite{Zocca99}).\\
As for mixing in radiative stellar interiors, it was suggested that the
discontinuity of molecular weight left by the 1st DUP could inhibit any
extra-mixing between the base of the convective envelope and the HBS. After
the {\em bump}, the $\mu$-gradients are much smoother in this region,
permitting some extra-mixing to occur (Sweigart \& Mengel \cite{SM79};
Charbonnel \cite{CC95}; Charbonnel et al. \cite{CBW98}).

\section{From the pioneering work on stellar rotation to the present 
treatment of the transport of angular momentum and chemicals}

Sweigart \& Mengel (\cite{SM79}, hereafter SM79) investigated the possibility that
meridional circulation might lead to the mixing of CNO-processed material in RGB stars.
Though the physics of rotation-induced mixing invoked at that time was very crude,
this pioneering work has magnificently settled the basis of a complex problem. 
SM79 discussed in great details the problem of $\mu$-gradients which
were known to inhibit meridional circulation (Mestel \cite{Mestel53}, \cite{Mestel57}).
For mixing to be efficient, they underlined the necessity for the radiative
zone separating the CE from the HBS not to present
significant molecular weight gradients.
The other crucial point made by SM79 concerned the importance of the
angular momentum (hereafter AM) distribution within the deep CE of
RGB stars on the resulting mixing of CNO processed material. Indeed, beyond
the 1st DUP, the angular velocity of
a radiative layer near the HBS depends sensitively on how much AM has been deposited
by the retreating CE. SM79 investigated
two extreme cases, namely a constant specific angular momentum  and a
uniform angular velocity within the CE.
Substantial CNO processing of the envelope could be obtained with plausible
MS angular velocity only when the inner part of the convection
envelope was allowed to depart from solid body rotation. As we shall see
in this paper, our lack of knowledge of the distribution of angular
momentum within the CE of giant stars remains one of
the weakest points of our global understanding of rotation-induced mixing at this phase.

Rotational transport processes cannot be simply reduced to meridional circulation.
Once established, this large scale circulation generates
advection of AM, and thus favours the development of various hydrodynamical
instabilities. 
Zahn (\cite{Zahn92}) proposed a description of the interaction between meridional circulation and
shear turbulence, pushing forward the idea of chocking the meridional circulation
by $\mu$-gradients.
Following these developments but using a simplified version of Zahn's description, 
Charbonnel (\cite{CC95}) re-investigated the influence of such a process 
in RGB stars.
She conjectured that the combination of the disappearance of the mean molecular weight
gradient barrier after the {\em bump} and the increase of mixing in the HBS has the
proper time dependence to
account for the observed behaviour
of carbon isotopic ratios and for the Li abundances in Population~II
low-mass giants.

In these exploratory computations however, the diffusion coefficient for chemicals was
 derived from an assumed constant rotation velocity (in the radiative zone)
 on the RGB, and the transport of AM by hydrodynamical processes was not
considered. This is however of utmost importance in the understanding of
the rotation-induced mixing (see the review by Maeder \& Meynet
\cite{MM00araa}).

Then, Denissenkov \& Tout (\cite{DT00}) applied the formalism of Maeder \&
Zahn (\cite{MZ98}) to a typical globular cluster RGB star. However, this
was done in a post-processing approach and thus, did not take into account
the feedback of mixing on the stellar structure.  Considering the angular
velocity at the base of the CE as a free adjustable parameter and treating
the transport of chemicals only beyond the {\em bump} as in the
aforementioned works, they obtained large diffusion coefficients able to
reproduce not only the Li, C and N abundance anomalies at the {\em bump},
but also the O-Na and Mg-Al anti-correlations. These two features being
clearly primordial, Denissenkov \& VandenBerg (\cite{DV03}) revised these
results. They simplified their previous approach, letting AM evolve only
due to structural readjustments (no rotational transport), and derived a
diffusion coefficient to be applied to the chemicals beyond the {\em
bump}. They considered the obtained mixing rate to have ``{\em the correct
order of magnitude}'', even though it is too low by a factor of 7 to
reproduce the observational data for Pop~I stars.

In the present paper we propose a self-consistent approach of
rotational-mixing in low-mass RGB stars. We define here as
self-consistent a model in which the transport of angular momentum
and of chemicals is coupled to the evolution of the star from the zero age
main sequence on. If meridional circulation and shear-induced turbulence
are the only transport processes of AM considered, the assumptions made on
the rotation profile concern solely the initial condition, i.e. the
rotation profile at the ZAMS (assumed to be uniform) and the rotation
regime in the convective envelope. At each evolutionary step we thus
compute the new rotation profile together with the associated transport
coefficients resulting from structural readjustments and transport
processes associated with rotation. The abundance profile of each chemical
is then modified under the effect of both mixing and nuclear reactions.  In
such a procedure, the stellar structure ``reacts'' to rotational-mixing.\\
We discuss the effects of rotation in RGB stars taking into account the
latest prescriptions for the transport of AM in stellar interiors, and the
associated mixing processes. We describe the physical inputs of our models
in \S~\ref{sec:phys} and \S~\ref{sec:numsim}, and their effects on the
angular velocity profiles and diffusion coefficients in
\S~\ref{sec:mix}. In \S~\ref{sec:signmix} we present the results for
structural evolution and surface abundance variations of rotating low-mass
Pop II stars from the ZAMS to the upper RGB. We then give a summary of our main results, and
discuss them in relation with previous works in
\S~\ref{sec:sumprevwork}, and propose new investigation paths in \S~\ref{sec:conclusion}.

\section{Physical inputs for the evolution of rotating stars}\label{sec:phys}

\subsection{Standard input physics}\label{subsec:inputs}
 
The models presented here were computed with STAREVOL V2.30, and the reader
is referred to Siess \etal (\cite{Siess2000}) and Palacios \etal
(\cite{PTCF03}) for a detailed description.
 Let us recall the main inputs.

The nuclear reaction rates have been updated using the version 5.0 of the
nuclear network generator NetGen available at IAA
{\tt (http://astropc0.ulb.ac.be/Netgen)}. By default the adopted rates are :
NACRE (Angulo \etal \cite{Ang99}) for charged particles, Bao \etal (\cite{Bao}) for
neutron capture rates, Horiguchi \etal (\cite{Ho96}) for experimental beta decay
rates and Caughlan \& Fowler (\cite{CF88}) otherwise.

For the radiative opacities, we use the OPAL tables\footnote{The OPAL tables
  used are based on a solar-scaled chemical mixture with possible
  enhancement of C and O, but do not include $alpha$-elements enhancement.} above 8000~K (Iglesias
\& Rogers \cite{IR96}) and at lower temperatures the atomic and molecular
opacities of Alexander \& Fergusson (\cite{Alex}). The conductive opacities
are computed from a modified version of the Iben (\cite{Iben75}) fits to
the Hubbard \& Lampe (\cite{Hu69}) tables for non-relativistic electrons and from Itoh
\etal (\cite{Ito83}) and Mitake \etal (\cite{Mi84}) for relativistic electrons.

The equation of state is described in detail in Siess \etal (2000) and
accounts for the non ideal effects due to coulomb interactions and pressure
ionization. The standard mixing length theory is used to model convection
with $\alpha_{\rm MLT} = 1.75$ and the atmosphere is treated in the gray
approximation and integrated up to an optical depth $\tau \simeq 5\times
10^{-3}$. 

\subsection{Transport of angular momentum}\label{subsec:AM}

The evolution of AM and chemical species
follow Zahn (\cite{Zahn92}) and Maeder \& Zahn
(\cite{MZ98}). Meridional circulation and turbulence induced by the secular
shear
instability are the two 
transport mechanisms considered here.
Within this framework, the transport of AM obeys an
advection/diffusion equation
\begin{equation}
 \rho \frac{{\rm d} \left( r^{2}\Omega \right) }{{\rm d}
   t}=\frac{1}{5r^{2}}\frac{\partial }{\partial r}\left( \rho r^{4}\Omega
   U_r \right) +\frac{1}{r^{2}}\frac{\partial }{\partial r}\left( r^{4}\rho \nu_v \frac{\partial \Omega}{\partial r} \right),
\label{momevol}
\end{equation}
where $\rho$, $r$ and ${\rm \Omega}$ have their usual meaning.
$\nu_v$ is the vertical component of the turbulent viscosity associated 
with the shear instability. $U_r$ is the vertical component of the meridional circulation
velocity, which, assuming shellular rotation\footnote{The condition of
shellular rotation is satisfied when turbulence is highly anisotropic 
and ensures $\Omega \approx
\Omega(P)$. In that case,
Zahn's formalism may be applied strictly. Otherwise, it represents a
first order approximation.}, is given by:
\begin{equation}
U_r = \frac{P}{C_{p}\rho T g [\nabla_{\rm ad}-\nabla + \varphi/\delta
  \nabla_{\mu}]} \times \left[ \frac{L}{M_*}(E_{\Omega }+E_{\mu})\right] 
\label{vcirc}
\end{equation}
where  $M_* = M(1 - \frac{\Omega^2}{2\pi G \rho_m})$. $E_{\Omega }$ and $E_{\mu}$ 
depend respectively on the relative
horizontal variation of the density, $\Theta = \frac{\tilde{\rho}}{\rho} = \frac{1}{3} \frac{r^2}{g}
\frac{\diff \Omega^2}{\diff r}$, 
and of the mean
molecular weight, $\Lambda = \frac{\tilde{\mu}}{\mu}$.
Detailed expressions for these quantities are given in Appendix A.
The evolution of $\Lambda$ depends on the competition between the vertical advection of a mean
molecular weight gradient $\frac{\partial \ln\overline{\mu}}{\partial r}$ and
its destruction by horizontal diffusion $D_h$ (see \S~\ref{subsec:cdiff})
 \begin{equation}
 \frac{\partial \Lambda}{\partial t} + U_r \frac{\partial
   \ln\overline{\mu}}{\partial r} = - \frac{6}{r^2}D_h \Lambda 
 \label{lambda}
 \end{equation}
This equation is obtained under the assumption that
\begin{equation}
\frac{D_v}{\ell _v^2} \ll \frac{D_h}{\ell _h^2}.
\end{equation}
$D_v$ is the vertical turbulent diffusion coefficient (see \S~\ref{subsec:cdiff}), $\ell _v$ and $\ell _h$ are the characteristic distance scales in the
vertical and horizontal directions respectively.
As in paper I, the transport of AM is computed 
by solving 5 first order differential equations 
(Eq.~\ref{lambda} + 4 equations resulting from the
splitting of Eq.~\ref{momevol})
with a Newton-Raphson relaxation method.
The upper boundary condition on $\Lambda$ has
however been modified compared to paper~I, and we used the continuity equation at the base
of the CE rather than assuming $\Lambda = 0$.

Effects of $\mu$--currents ($E_\mu$ term in Eq.~\ref{vcirc}) are taken into account consistently from the
ZAMS up to the upper RGB.

\subsection{Transport of chemicals}\label{subsec:chem}

In the presence of strong anisotropic turbulence, Chaboyer \&
Zahn (\cite{CZ92}) showed that 
the vertical advection of chemicals by a large scale circulation combined with
strong horizontal diffusion produces a vertical effective diffusivity $D_{\rm eff}$ (see
\S~\ref{subsec:cdiff}).
The vertical transport of a chemical species $i$ of
concentration $c_i$ can thus be described by a 
pure diffusion equation:
 \begin{equation}
\centering{
\rho  \frac{{\rm d} {c_i}}{{\rm d} t} = \underbrace{\dot{c_i}}_{\rm \it nuclear} +
\underbrace{\frac{1}{r^2}\frac{\partial}{\partial r}\left[r^2\rho
   U_{ip} {c_i}\right]}_{\rm \it atomic~diffusion} +
\underbrace{\frac{1}{r^2}\frac{\partial}{\partial r}\left[r^2\rho
   D_{\rm tot} \frac{\partial {c_i}}{\partial
     r}\right]}_{\rm \it macroscopic~processes}.
}
\label{diffelts}
\end{equation}
$U_{ip}$ is the atomic diffusion velocity of the element with respect to protons, and
$D_{\rm tot}$ is the total macroscopic diffusion coefficient, and is the sum
of the effective diffusion coefficient (\ref{deff}) and of the vertical turbulent
diffusion coefficient $D_v$ (see \S~\ref{subsec:cdiff}).

The diffusion equation (\ref{diffelts}) is solved for each of the 53
species considered in the code considering $\dot{c_i} = 0$ (that is, diffusion and
nucleosynthesis are decoupled). Here again, we used a Newton-Raphson method, as for the
structure and AM transport equations.

\subsection{Diffusion and viscosity}\label{subsec:cdiff}

Let us briefly recall the various formulations used for the diffusion
coefficients entering Eqs.~(\ref{momevol}), (\ref{lambda}) and
(\ref{diffelts}).
\begin{itemize}

\item[$\bullet$] $\nu_v$, $D_{v}$\\ As in paper~I, we assume that the
secular shear instability dominates and that vertical shear eventually
becomes turbulent in the radiative stellar interiors due to the low
viscosity of the plasma (Zahn \cite{Zahn74}). The development of turbulence 
is subject to the Reynolds criterion and sets in when
$$ Re = \frac{\nu_v}{\nu_{\rm mol} + \nu_{\rm rad}} > Re_c,$$ where 
$\nu_{\rm rad} = \frac{4}{15}\frac{aT^4}{\kappa c \rho^2}$ is the radiative 
viscosity, $\nu_{\rm mol}$ is
the molecular viscosity and $Re_c \simeq 10$ is the critical Reynolds number. The shear
instability obeys the Richardson criterion, and according to the classical
formulation should set in when
$$ \frac{\diff u}{\diff z} > \frac{N^2}{Ri_c},$$ where $Ri_c = 0.25$ is
the critical Richardson number. Here, we rather consider a
modified Richardson criterion to take into account radiative losses and/or
horizontal diffusion as described below. Several modified criteria have
been proposed; in this paper we will compare results obtained with two of
them.

\begin{itemize}
\item The first criterion assumes that thermal diffusion reduces the stabilising
effect of thermal stratification without affecting the chemical
part. It leads to (cf. Maeder \& Meynet
\cite{MM96}):
\begin{equation}
 D_v = \nu_v = \frac{8}{5} K_T \frac {Ri_c  (r d
 \Omega/\diff r)^2 - N^2_{\mu}}{N^2_T},
\label{eqDvMM96}
\end{equation}
where $N^2 = N^{2}_{T} + N^{2}_{\mu}$ is the 
Brunt-V\"ais\"a\"al\"a frequency and $K_T$ is the thermal diffusivity.
In this paper we will refer to this prescription as 
\emph{MM96}.
\item The second criterion, which we used in paper I, also considers
the erosion of the chemical stratification by the large horizontal diffusion
(cf. Talon \& Zahn \cite{TZ97}):
\begin{equation}
 D_v = \nu_v = \frac{8}{5} \frac {Ri_c  (r d
 \Omega/\diff r)^2}{N^{2}_{T}/(K_T+D_h)+N^{2}_{\mu}/D_h},
\label{eqDvTZ97}
\end{equation}
where $D_h$ is the horizontal
turbulent viscosity. 
In this paper we will refer to this prescription as 
\emph{TZ97}.
\end{itemize}
Let us further note that, due to the lack of a better prescription, we assume $D_v = \nu_v$.

\item[$\bullet$] $D_{\rm eff}$\\ The effective diffusion coefficient solely
appears in the transport equation of chemicals. In the approximation
of highly anisotropic turbulence, it is the diffusive representation of the
effects of meridional circulation (Chaboyer \& Zahn \cite{CZ92}), and can
be written as follows :
\begin{equation}
D_{\rm eff} = \frac{|r U_r|^2}{30 D_h} 
\label{deff}
\end{equation}

\item[$\bullet$] $\nu_h$, $D_h$\\
The vertical turbulent diffusion coefficient $D_v$ as well as the effective
diffusion coefficient $D_{\rm eff}$ depend on the horizontal component of
the turbulent diffusivity $D_h$. No description of this diffusivity can be
drawn from first principles, and its expression has to be parametrised.
Assuming that the differential rotation on an isobar is small compared to
unity, Zahn (\cite{Zahn92}) first proposed the following expression for
$\nu_h$: 
\begin{equation}
\nu_h = \frac{r}{C_h}\left|\frac{1}{3 \rho r}\frac{{\rm d} (\rho r^2 U)}{{\rm d}
    r}-\frac{U}{2}\frac{{\rm d} \ln r^2\Omega}{{\rm d} \ln r}\right| \equiv
    \frac{r}{C_h}\left| 2 V - \alpha U \right| ,
\label{Dh}
\end{equation}
where $C_h$ is a free parameter of order 1
which we used in our paper I, $V = \frac{1}{6 \rho r}\frac{\rm d}{{\rm d}r}\lp \rho r^2 U \rp$ is the
horizontal component of the meridional velocity and $\alpha = \frac{1}{2}
\frac{{\rm d} \ln r^2 \bar{\Omega}}{{\rm d} \ln r}$. Meanwhile, some improvements have been
achieved and more realistic prescriptions for the horizontal shear
turbulent diffusivity including a dependence on the rotation rate were
derived by Maeder (\cite{Maeder03}) and Mathis \etal (\cite{MPZ04}) :
\begin{equation}
\nu_h = \chi^{\frac{1}{n}}~r~ \lp r~\bar{\Omega}(r)~V^k~[2V - \alpha U] \rp ^{\frac{1}{n}},
\label{Dhnew}
\end{equation}
where  ($\chi$~;~$n$~;~$k$) = ($\frac{3}{400 m
  \pi}$~;~3~;~1) in Maeder's expression (with $m$ = 1,3 or 5), and ($1.6
10^{-6}$~;~2~;~0) in Mathis \etal (\cite{MPZ04}).
In Maeder (\cite{Maeder03}), the horizontal turbulent viscosity $\nu_h$ is
derived from a comparison between the dissipation rate of turbulent energy
by meridional circulation and the viscous dissipation rate, while in Mathis
et al. (\cite{MPZ04}), it is derived from Couette-Taylor laboratory
experiments (see Richard \& Zahn \cite{RZ99}).\\
The horizontal turbulent diffusion coefficient $D_h$ obtained by
Eq.~(\ref{Dhnew}) is larger than the one derived from expression
Eq.~(\ref{Dh}), and is more consistent with the shellular rotation
hypothesis. 
In the following, we will compare results obtained with these two
  prescriptions, which we refer to as $Zahn92$ and $MPZ04$ (see Table~\ref{table1}).
\end{itemize} 

\subsection{Rotation law in the convective envelope}\label{subsec:rotCE}

The formalism developed by Zahn (\cite{Zahn92}) 
describes the transport of AM in
radiative zones, and the rotation profile in the CE is defined by an
upper boundary condition.
Its choice is however of prime importance since it determines the flux of angular
momentum between these two regions.\\ 
In the case of RGB stars, this will play an essential role as discussed in \S 5.1.

The interaction between rotation and
convection is a longstanding and not fully understood problem, and despite
the development of 3D numerical simulations (Ballot \etal \cite{Ballot04};
Browning \etal \cite{Browning04}), the rotation profile 
within deep convective envelopes
remains unknown. As already suggested by SM79, we may consider two limiting cases for the CE rotation law:
\begin{enumerate}

\item {\it Uniform angular velocity (solid body rotation)}\\ This
hypothesis has generally been assumed when modelling the evolution of
rotating stars (Endal \& Sofia \cite{ES78}; Talon \etal \cite{TZMM97};
Meynet \& Maeder \cite{MM00}; paper I).  It is also the rotation regime
obtained when describing AM transport in the CE by a
diffusion equation using the diffusion coefficient derived from the MLT
theory (Heger \etal \cite{HLW00}). Imposing solid body rotation in the CE
is thus equivalent to assuming that the meridional currents are inhibited in the
presence of convection, and that the turbulent viscosity
associated with convection is large enough to allow for instantaneous
homogenisation of the angular velocity profile $\Omega(r)$, as it is the
case for chemicals (Endal \& Sofia \cite{ES76}).  This condition is also
motivated by observations of the solar convection zone (the sole star
for which we have such informations at present), where {\em radial}
differential rotation is minute (Kosovichev \etal \cite{kosovichev97}).

\item {\it Uniform specific angular momentum (differential rotation)}\\
Already in the early 70's, Tayler (\cite{Tayler73}) addressed the effects
of rotation in stellar convective zones and came to the conclusion that
meridional currents could develop and alter the rotation law in stellar
convective regions. With caution, he suggested that ``{\it it is possible
that the asymptotic state [of rotation in a convective zone] is closer to
one of uniform angular momentum than uniform angular velocity}''. Some
years later, Sweigart \& Mengel (\cite{SM79}) proposed that
rotation-induced mixing by meridional circulation could explain both the
CNO abundance anomalies in RGB stars and the slow rotation rates observed
in MS low-mass stars provided the radiative interior conserves its angular
momentum during the first ascent of the giant branch, and the CE has
\emph{constant} 
($\diff j_{\rm CE}/\diff t = 0$) 
and \emph{uniform} ($\diff j_{\rm
CE}/\diff r = {\rm cst}$) specific angular momentum. More recently, Denissenkov \&
Tout (\cite{DT00}) investigated both possibilities for rotation in the
CE of a low-mass RGB star, and concluded in favour of uniform and constant
specific angular momentum in the CE.  Investigating the rotation rates on
horizontal branch (HB) stars of globular clusters,
Sills \& Pinsonneault
(\cite{SP00}) showed that slow rotation on MS stars and $\upsilon \sin i$
up to $40~\kms$ on the HB indicate that a
non-negligible amount of
AM is preserved in the stellar interior between the MS
turn-off and the HB. They proposed that this could be achieved assuming
uniform specific angular momentum in the CE during the RGB phase, a
conclusion again similar to that drawn by SM79 and more recently, by
Chanam\'e \etal (2004a, b).

These studies indicate that the condition of differential rotation in the
CE (i.e. $j_{\rm CE}(r) = \frac{2}{3} r^2 \Omega (r) = {\rm cst}$) could be
a key ingredient to derive a consistent history of the AM
evolution in low-mass stars, as well as to produce a high degree of mixing
in the RGB interiors needed to explain part of the abundance anomalies
observed at this phase.

\end{enumerate}
\begin{table}
\caption{
Rotation input physics of the models.
  \emph{$D_h$} : \emph{MPZ04} refers to Mathis et al. (2004) whereas \emph{Zahn92}
  refers to Zahn (1992). \emph{$D_v$} : \emph{TZ97} refers to Talon \& Zahn (1997) and \emph{MM96}
  refers to Maeder \& Meynet (1996). Uniform specific angular momentum in the CE was
  only applied beyond the turn-off. All models were computed with uniform
  angular velocity in the CE during the MS phase. Braking, if applied, begins at the
  ZAMS following a Kawaler (1988) law calibrated in order to lead to
  equatorial rotational velocity of $\upsilon \approx 10 \kms$ when they
  reach the Hyades age (this occurs for 
  $T_{\rm eff} \approx 6300~{\rm K}$)
  early on the main sequence. This velocity is typical of cool stars in
  the Hyades.
  The highlighted row indicates our reference model M1.}
 \begin{tabular}{c | c | c | c | c | c | c}
 \hline
 \hline
 {\scriptsize Model} & {\scriptsize CE} & {\scriptsize $\upsilon_{\rm
 ZAMS}$} & {\scriptsize Braking} & {\scriptsize $D_h$} & {\scriptsize
 $\mu$-currents} & {\scriptsize $D_v$}\\
 & {\scriptsize rotation law} & {\scriptsize $\kms$} & & & &\\\hline
 M0 & --- & 0  & --- & --- & --- & ---\\
\rowcolor{yellow} {\bf M1} & {\bf $\Omega (r) = $ {\scriptsize cst}} & {\bf 5} & { \bf \scriptsize no} &
 {\scriptsize \bf MPZ04} & yes & {\scriptsize TZ97}  \\
 M2 & $ j(r) = $ {\scriptsize cst}  & 5  & {\scriptsize no} &
 {\scriptsize MPZ04} & yes & {\scriptsize TZ97} \\
 M3 & $ j(r) = $ {\scriptsize cst} & 5  & {\scriptsize no} &
 {\scriptsize MPZ04} & no & {\scriptsize TZ97} \\
 M4 & $ j(r) = $ {\scriptsize cst} & 5  & {\scriptsize no} &
 {\scriptsize MPZ04} & yes & {\scriptsize MM96} \\
 M5 & $ j(r) = $ {\scriptsize cst} & 5 & {\scriptsize no} &
 {\scriptsize Zahn92} & yes & {\scriptsize TZ97} \\
 M6 & $ j(r) = $ {\scriptsize cst} & 110 & {\scriptsize yes} &
 {\scriptsize MPZ04} & yes & {\scriptsize TZ97} \\\hline \hline
\end{tabular}
\label{table1}
\end{table}

\begin{figure*}
\begin{center}
\includegraphics[width=12cm]{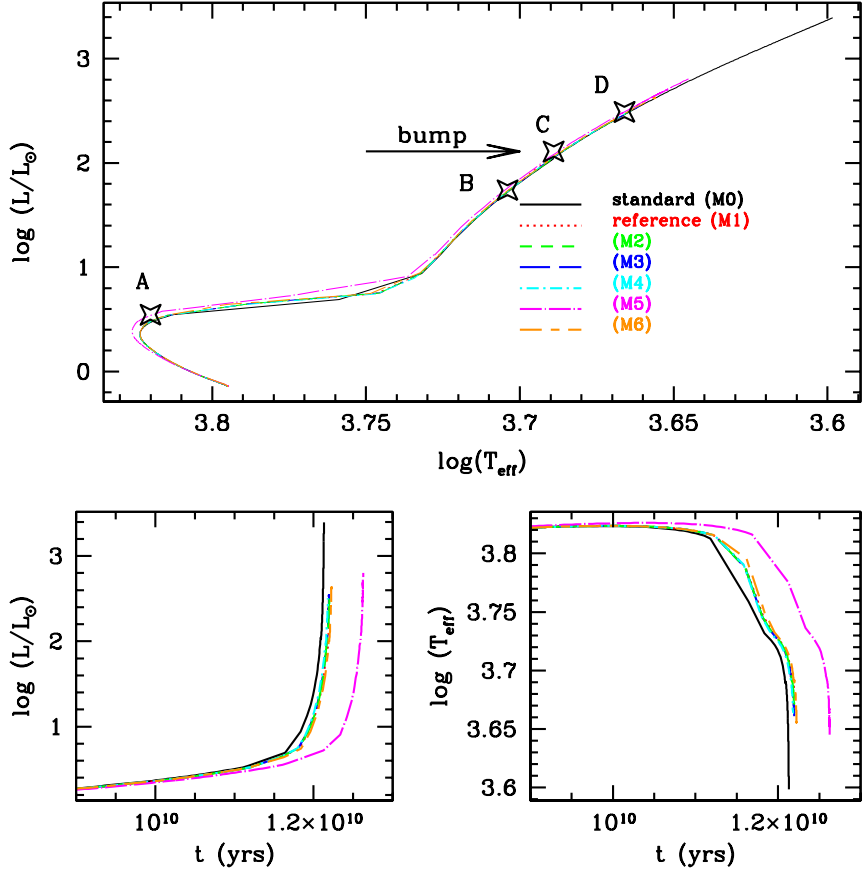}
\caption{ {\em \,Upper panel}: Hertzsprung-Russell diagram for the
  different test models presented. All sequences start with $M_{\rm ZAMS} = 0.85 \,{\rm
  M}_{\odot}$, [Fe/H] = $-1.57$. Evolutionary tracks for models M1 and
  M3 are superimposed. Open symbols labelled A, B, C and D correspond
  to the turn-off, the end of the 1st DUP, the {\em bump}
  and $L = 310 \,{\rm L}_\odot$ respectively.{\em \,Lower panels}: Evolution of luminosity
  (left) and effective temperature (right) as a function of time beyond the
  turn-off.}
\end{center}
\label{HRD}
\end{figure*}

\begin{figure*}[ht]
\begin{center}
\includegraphics[height=15cm,width=12cm,angle=-90]{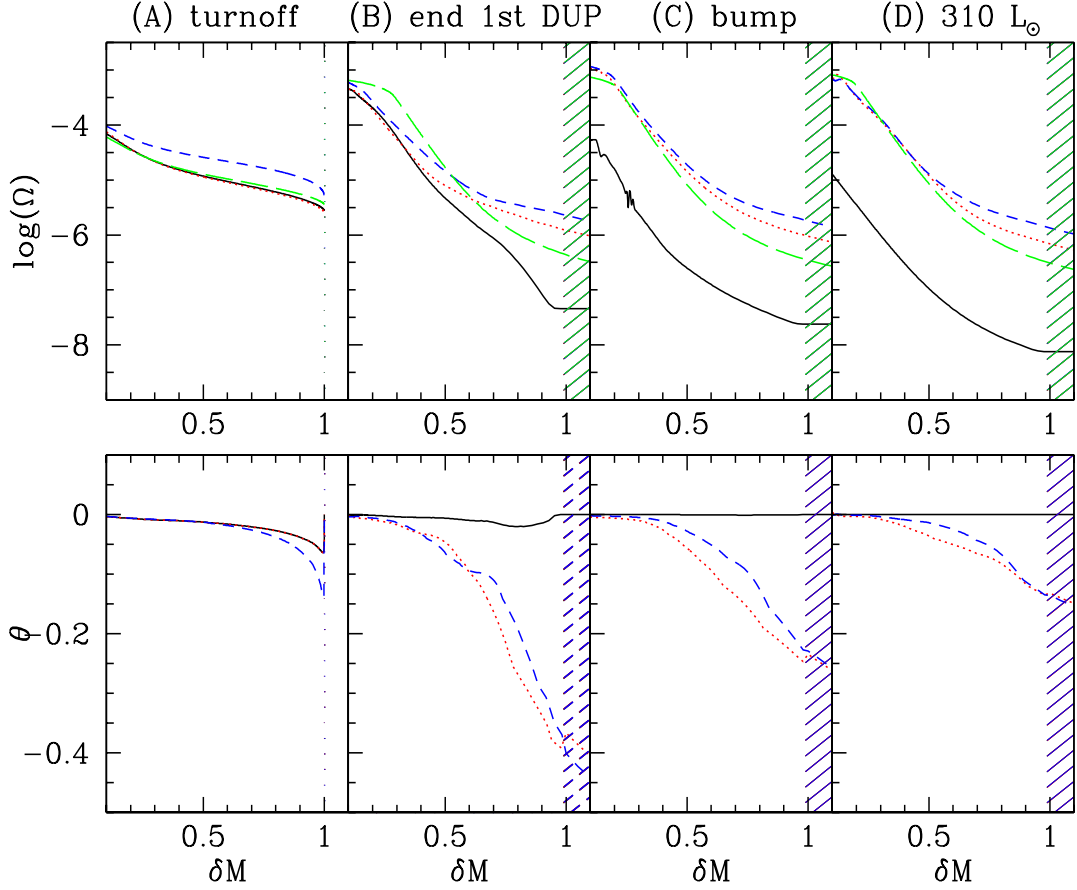}
\end{center}
\caption{Angular velocity $\Omega$ and horizontal density fluctuations $\theta \propto
  \partial \Omega / \partial r$ inside models
  M1 (solid lines), M2 (dotted lines) and M6 (dashed lines) at the four
  evolutionary points top-labelled on the graph and reported on
  Fig~\ref{HRD}. $\Omega$ and $\theta$ are
  plotted against the scaled mass coordinate $\delta M$ which allows a
  blow-up of the region of interest ($\delta M = 0$ at the base of the HBS,
   and $\delta M = 1$ at the base of the CE). The long dashed lines on the
  upper pannels represent the $\Omega$ profiles that one gets when the
  angular momentum evolution in model M2 only results from the structural
  readjustments beyond the turn-off (i.e. no AM transport).
}
\label{omegatheta}
\end{figure*}

\begin{figure*}[ht]
\begin{center}
\includegraphics[height=15cm,width=12cm,angle=-90]{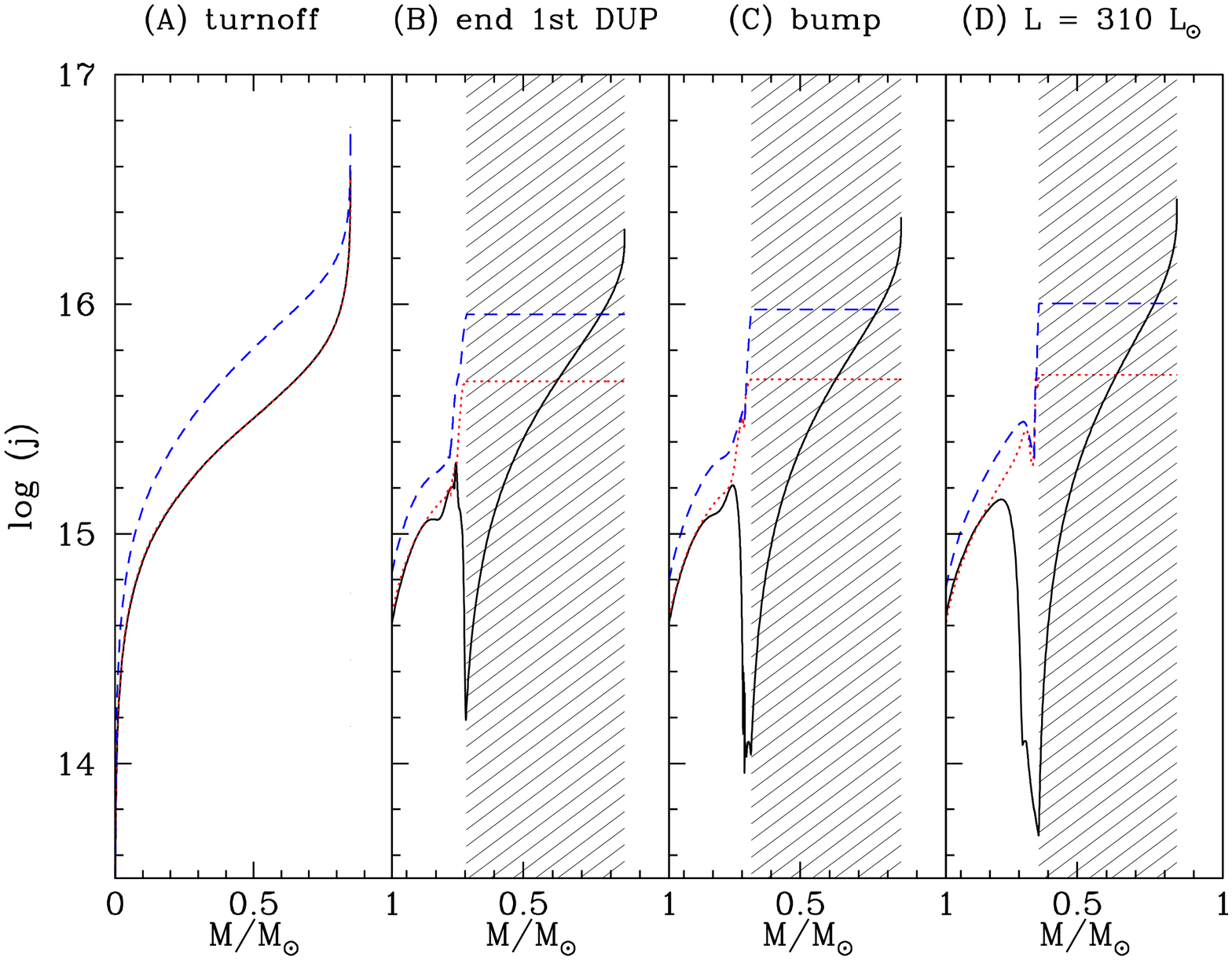}
\end{center}
\caption{ Profile of the logarithm of the specific angular momentum $\log (j)$ inside models
  M1 (solid lines), M2 (dotted lines) and M6 (dashed lines) at the four
  evolutionary points top-labelled on the graph and reported on
  Fig.~\ref{HRD}, as a function of the internal mass coordinate
  $M_r$ in units of \msun . 
}
\label{momspec}
\end{figure*}

\section{Numerical simulations}\label{sec:numsim}

We model a typical metal-poor globular cluster RGB star, with initial mass $M =
0.85$\,\msun, initial helium content $Y_{\rm
  ini} = 0.248$ and $[{\rm Fe/H}] = -1.57$
which corresponds to the metallicity of M13 stars (Sneden \etal \cite{SKGPF04}).
We take into account the
$\alpha$-enrichment expected at such metallicities, with [O/Fe] =
[Ne/Fe] = [Mg/Fe] = $+0.3$ dex, and account for the odd-even effect on sodium, with
[Na/Fe] = $-0.3$ dex. The initial ratio for the magnesium isotopes is
$^{24}{\rm Mg}:^{25}{\rm Mg}:^{26}{\rm Mg} = 87.5:6:6.5$, and is similar to the values
determined by Yong et al. (\cite{Yong03})
for the stars with the lowest $^{25}{\rm Mg}$ and $^{26}{\rm Mg}$ in NGC 6752, a
globular cluster with [Fe/H]$ \approx -1.6$. The mass fractions of the other
elements are in solar system proportion, and scaled for their sum to be
equal to unity. This corresponds to a metallicity of $Z = 8~10^{-4}$. 

Mass loss is included from the ZAMS on in all our models. We use the empirical Reimers
(\cite{Reimers75}) formula with a metallicity scaling :
\begin{equation}
\dot{M} = -3.98 10^{-13} \eta \frac{L\,R}{M}\sqrt{\frac{Z}{Z_{\odot}}} ~~ 
{\rm M}_{\odot}\,{\rm yr}^{-1}
\end{equation}
with $\eta = 0.5$. At each time step, the associated AM losses
are taken into account and the adjusted total AM is
conserved when applying rotational transport.

All models are computed assuming uniform angular velocity in the CE during
the MS as indicated by the solar case. 
When considered, the hypothesis of uniform specific angular momentum in
the CE is applied beyond the turn-off, i.e. when the convective envelope begins to deepen.
We will compare two different
rotational histories, namely models that were slow rotators already on the ZAMS
and models with an initially larger velocity typical of ZAMS Pop~I stars, but
which experience magnetic braking in their early evolution, as expected from solar-type stars.  
This braking results in similar surface velocities at half-way on the main
sequence (when $X_c \approx 0.5$) for initially fast and slow rotators. 
Considering their important role in shaping the rotation profile 
(see paper~I), 
the so-called $\mu$-currents ($E_{\mu}$ term in
Eq.~\ref{vcirc}) are taken into account in all our rotating models except
M3. In all cases the initial rotation profile at the ZAMS is defined by
$\Omega(r) = \upsilon_{\rm ini} / R_{\star} = {\rm cst}~~\forall~~{\rm r}$.

We finally underline 
that Eq.~(\ref{momevol}) is solved in all its
complexity for all the rotating models presented here from the ZAMS up to
the upper RGB.

Table~\ref{table1} lists the characteristics of the models
that we have computed using different prescriptions for the input physics. 
For our reference model (M1), we consider the following set of
parameters/physical ingredients: uniform angular velocity in the convective
regions at all times, initial surface velocity of 5 $\kms$ on the ZAMS and
no braking applied, Mathis et al. (\cite{MPZ04}~; Eq.~\ref{Dhnew}) and Talon \& Zahn
(\cite{TZ97}~;~Eq.~\ref{eqDvTZ97}) prescriptions for the
horizontal $\nu_h$ and vertical $\nu_v$ turbulent diffusion coefficients respectively.

Figure~\ref{HRD} presents the Hertzsprung-Russell diagram and the evolution
of the surface luminosity and temperature for the models computed. It will
be discussed in more details in \S~\ref{subsec:structure}. The evolutionary points on which
we will focus in the following sections are marked on the evolutionary
path and correspond to the turn-off (A), the end of the first DUP (B), the
bump (C) and $L = 310 \,{\rm L}_\odot$ (D).

\section{Testing the physics of rotation}\label{sec:mix}

In this section, we analyse the impact of the different physical inputs included in
the models presented in Table~\ref{table1}.

\subsection{Rotation in the convective envelope}\label{subsec:rotenv}

In \S~\ref{subsec:rotCE} 
we mentioned the uncertainty regarding the rotation
law in the convective envelope of a giant star. Previous studies also
provide some hints that the
evolution of the AM (and chemicals) distribution within the
radiative interior might strongly depend on the rotation regime 
at the base of the 
convective envelope. This aspect can be studied by comparing models M1 and
M2, which solely differ by the applied rotation law in the CE beyond the turn-off.

\begin{figure}[t]
\resizebox{\hsize}{!}{\includegraphics{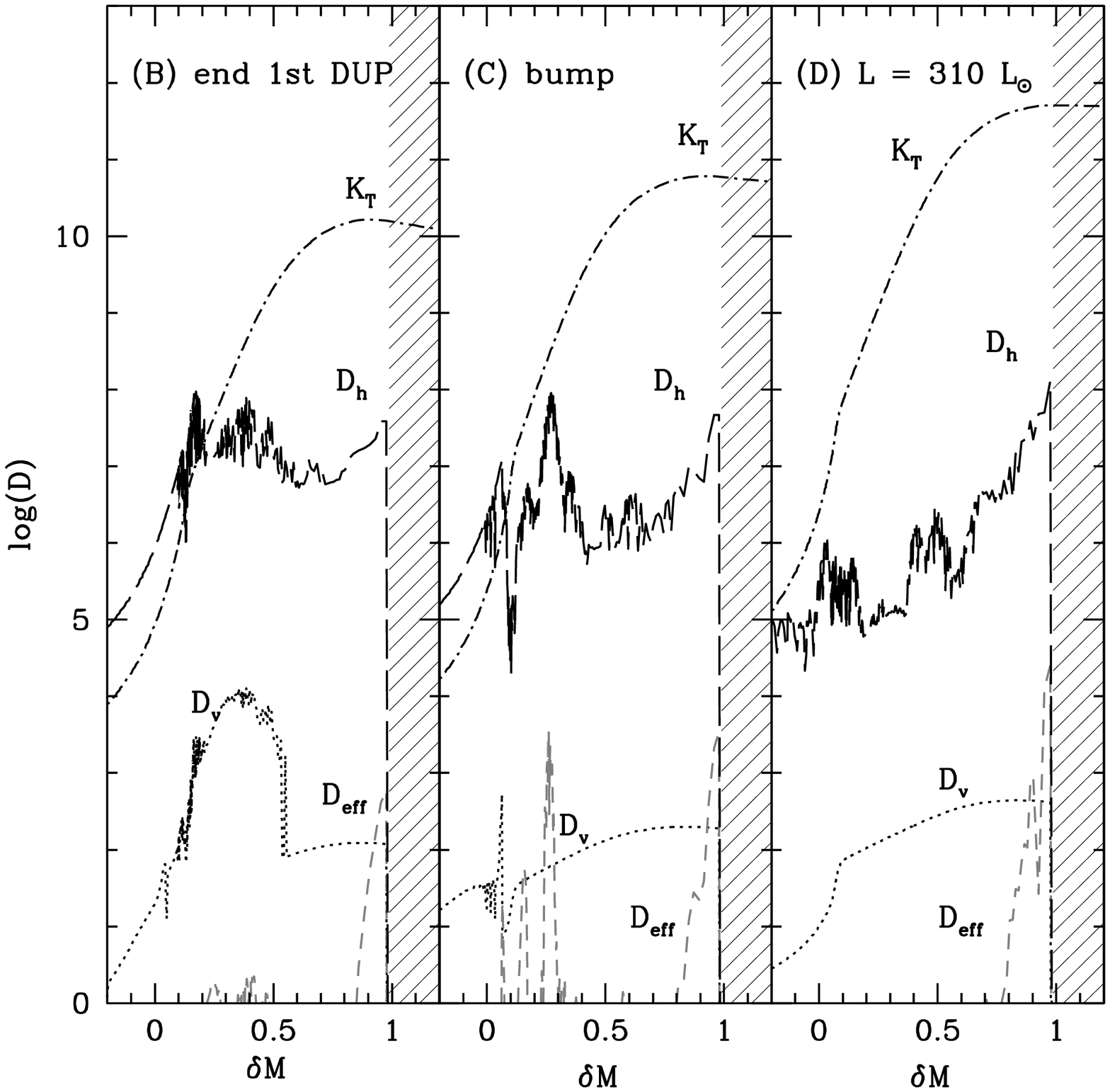}}
\caption{Profiles of the different diffusion coefficients at the end of the
  first dredge-up (left panel) and at the {\em bump} (right panel) for our
  reference model (M1 in Table~1).
}
\label{cdiffmodref}
\end{figure}

Profiles of the angular velocity $\Omega$, of the horizontal
density fluctuations $\theta$ ( = $\frac{\Theta}{\Omega} = \frac{1}{3} \frac{r^2}{g} \partial
\Omega/\partial r$) and of the specific angular momentum $j$ at four different
evolutionary stages for models M1, M2 and M6 are presented in Figs.~\ref{omegatheta}
and~\ref{momspec}. 
In Fig.~\ref{omegatheta}, as well as in several 
other figures in this paper, quantities are plotted against
$\delta M$ instead of $M_r$. $\delta M$ is a relative mass coordinate allowing for a
blow-up of the radiative region above the HBS, 
and is defined as
\begin{equation}
\delta M = \frac{M_r - M_{\rm HBS}}{M_{\rm BCE} - M_{\rm HBS}}.
\end{equation} $\delta M$ is
equal to 1 at the base of the convective envelope 
and 0 at the base of the HBS (where $X = 10^{-7}$). Typically in our
  models, the nuclear reactions occur between $\delta M$ =
  0.2 and $\delta M$ = 0.5, a large
mean molecular gradient being associated with the layers of maximum energy
production at $\delta M$ = 0.2

By assumption, the rotational evolution is identical in models M1 and M2 up to the turn-off. 
Beyond, the evolution of the AM distribution is dominated by structural readjustments of the 
star becoming a giant. 
The degree of
differential rotation in the radiative zone globally increases with time,
leading to a rapidly rotating core and a slowly rotating surface ($\Omega$
rises by 3 to 4 orders of magnitude between the base of the CE and the edge of the
degenerate He core in both models).\\ 
At the turn-off, the base of the CE rotates at the same velocity in models
M1 and M2. As the star crosses the Hertzsprung gap and approaches 
the Hayashi line on its way to the red giant branch, the core contracts 
while the outer layers expand and are thus
efficiently slowed down. For model M1, the
velocity of the inner shells of the envelope also decreases substantially
because of the solid body rotation of the CE ($\diff \Omega_{CE}/ \diff r = 0 $), and the
AM attached to the convective envelope concentrates in the
outer layers. At the end of the 1st DUP the angular velocity at the base of
the CE, $\Omega_{BCE}$, decreased by a factor of 65. When the CE withdraws
in mass, it leaves behind radiative shells with low specific
angular momentum that rotate slowly. The
resulting differential rotation rate is low in this region as 
shown in the $\theta$ and
$\Omega$ profiles.

 For model M2, during the 1st DUP the surface layers
slow down but the uniformity of the specific angular momentum 
prevents the shells of the inner CE from decelerating abruptly and ensures the
concentration of AM in this denser region. At the end of the 1st
DUP, $\Omega_{BCE}$ has decreased only by a factor of 3. As the CE withdraws in mass,
its deeper shells retaining the larger part of the CE angular momentum fall
back into the underlying radiative zone, where they conserve their large angular velocity and strong
differential rotation.

Figure~\ref{momspec} shows that after the completion of the 1st DUP,
  the assumption of uniform specific angular momentum within the CE also
  translates into a quasi-constancy of the level of specific angular
  momentum in this region (dotted and dashed lines in panels (B), (C) and
  (D)). This is a natural expectation since the convective envelope of a
  red giant represents more than 80 \% of the stellar radius and retains
  most of the angular momentum. The assumption of {\em constancy} of
  $j_{CE}$ under a regime of uniform specific angular momentum remains
  however an approximation since transfer of AM occurs between the interior
  and the envelope. Our models indicate that the specific angular momentum
  in the envelope indeed slightly increases with time.

The difference of the rotation profiles encountered in models M1
and M2 below the CE affects the diffusion coefficients, in particular $D_v$, which
scales as $(r \diff \Omega/ \diff r)^2$.

\begin{figure}[t]
\begin{center}
\resizebox{\hsize}{!}{\includegraphics{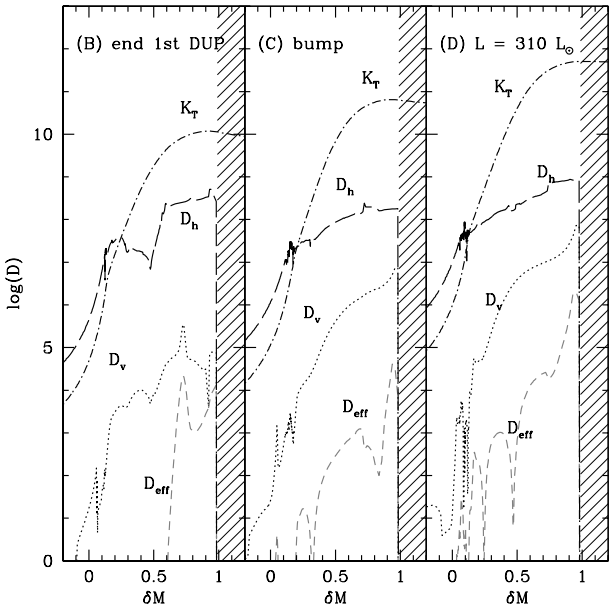}}
\end{center}
\caption{Same as Fig.~\ref{cdiffmodref} for M2 in Table~1. M2
  only differs from M1 by the adopted differential rotation law in the convective envelope}
\label{cdiffMod2}
\end{figure}

Figures~\ref{cdiffmodref} and \ref{cdiffMod2} present the profiles of the
diffusion coefficients entering the transport equations of AM
and chemicals 
(Eqs.~\ref{momevol} and~\ref{diffelts}) 
at the end of the 1st
DUP, at the {\em bump} and at $L = 310\,\Lsun$ for models M1 and M2, respectively. One may 
notice that $D_h$ is everywhere 
much larger than $D_v$ (see also Figs.~\ref{cdiffMod5},~\ref{cdiffMod6} and~\ref{cdiffMod4}). This
validates the shellular rotation hypothesis.

\begin{itemize}
\item \emph {Model M1} \\ At the end of the 1st DUP (left panel in
Fig.~\ref{cdiffmodref}), differential rotation is negligible below the CE
in the region between $\delta M \simeq 0.5$ and $\delta M$ = 0.91. There,
the vertical turbulent viscosity is smaller than its critical value given
by $\nu_{v,\rm crit} = Re_c \nu_{\rm rad} \simeq 3000$, and turbulence does
not develop. \emph{The low rate of differential rotation quenches the
turbulent transport and creates a gap that disconnects the CE from the
regions where nucleosynthesis occurs.}  Below $\delta M$ $\simeq 0.5$ the
differential rotation is larger and the shear turbulence dominates the
transport down to the region ($\delta M \approx 0.1$) where both meridional
circulation and shear turbulence are very efficiently hindered by the high
mean molecular weight barrier associated with the hydrogen burning (see
upper panels of Fig.~\ref{cdiffall}). At the {\em bump} and beyond,
shear-induced turbulence does not develop and the effective diffusion
coefficient remains negligible below $\delta M \simeq 0.85$, so that no
modification of the surface abundance pattern by rotational mixing should
be expected beyond this point (see \S~\ref{subsec:abund}).

\item \emph {Model M2} \\ When assuming uniform specific angular momentum
in the CE, the picture is very different (Fig.~\ref{cdiffMod2}). The
larger differential rotation rate allows the shear flow to become
turbulent almost everywhere between the base of the CE and the HBS. Shear
turbulence dominates the transport of chemicals across the entire radiative
zone, the effective diffusion coefficient associated with meridional
circulation being always much smaller in the whole radiative region. 
Anticipating the results of
\S~\ref{subsec:abund}, we can already see from Fig.~\ref{cdiffall} that
$D_{\rm tot}$ does not rise above $10^5 {\rm cm}^2.{\rm s}^{-1}$ in the
outer HBS (around $\delta M$ = 0.2-0.3) even at the {\em bump}, which
is much lower than the parametric ``canonical mixing rate'' of $4.10^8 {\rm
cm}^2.{\rm s}^{-1}$ deduced from observational constraints (Denissenkov \&
VandenBerg \cite{DV03}).

\end{itemize}

The profile of $D_{\rm eff}$ in Fig~\ref{cdiffMod2} appears to be
  quite ragged below $\delta M \simeq 0.5$, in particular on panel
  (D). Each of the bump in this region is associated with an inversion of
  the meridional circulation velocity, which can be positive or
  negative. Although the meridional circulation can actually present
  various cells, these particular features are due to numerical
  instabilities occurring in the regions where the mean molecular weight
  gradients are non-negligible. When the effects of $\mu$-currents are
  included (terms depending on $\mu$ or $\Lambda$ in Eq.~\ref{vcirc}), the
  numerical system is highly non-linear and may be sensitive to numerical
  parameters such as the spatial and temporal resolutions.\\ The global
  amplitude of $D_{\rm eff}$ (and $D_v$) in these regions of large
  $mu$-gradients, together with the decrease in the diffusion coefficients
  in the nuclearly active shells are however robust results. Just below the
  CE, the first bump in the $D_{\rm eff}$ profile is also a robust feature,
  associated with the Gratton-\"Opik meridional circulation cell. These remarks apply to all
  our models.

\subsection{Impact of the ZAMS rotation velocity}\label{initrot}

We investigated the impact of the initial rotational
velocity considering an originally slow (model M2) and a fast
rotator (model M6) on the ZAMS, without changing any of the other
physical parameters. We have applied to model M6 the same treatment as to
Pop I stars (see paper I), namely a strong magnetic braking according to the
Kawaler (\cite{Kawaler88}) prescription 
so as to get a surface equatorial velocity lower than 10~$\kms$ before the central
hydrogen mass fractions gets lower than 0.5. At the turn-off the surface velocities of models M6 and M2 are
thus quite similar (see Table~\ref{table2}).\\ The point under scrutiny in this section is 
to determine whether this strong braking, which triggers strong turbulence in
the radiative zone during the main sequence, has an incidence on the
angular momentum distribution and on the diffusion coefficients beyond the
turn-off.\\
In the present study, we did not investigate the case of fast rotators at the
turn-off, this configuration being ruled out by the velocities available from observations of
low-mass turn-off stars in globular clusters (Lucatello \& Gratton
\cite{LG03}).

\begin{figure}
\resizebox{\hsize}{!}{\includegraphics{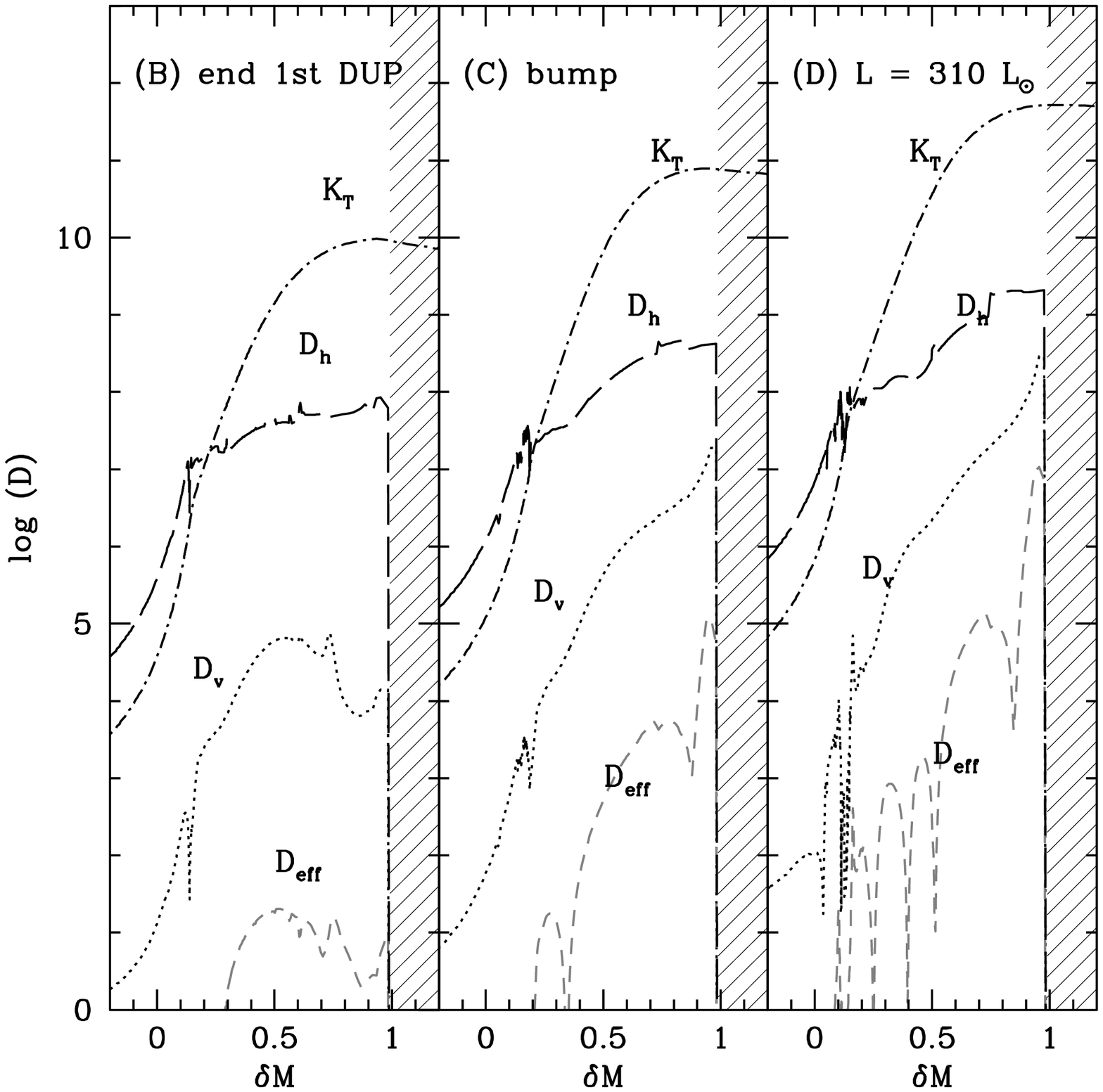}}
\caption{Same as Fig.~\ref{cdiffmodref} for M6. In this
  model, $\upsilon_{\rm ZAMS} = 110~{\rm km}.{\rm s}^{-1}$ and after
  efficient braking through according to a Kawaler law , $\upsilon_{\rm TO} = 6.4~{\rm
  km}.{\rm s}^{-1}$}. Uniform specific angular momentum was assumed in the CE.
\label{cdiffMod5}
\end{figure}

Figure~\ref{omegatheta} presents the degree of differential rotation
($\theta$ profiles) and the angular velocity inside models M2 and M6
(dotted and dashed lines respectively) at different evolutionary points.
Model M6 rotates globally faster than model M2 at the turn-off, and
this difference is maintained during the evolution. Indeed, although
model M6 undergoes a very efficient braking on the early MS, it is only
braked down to 5.7~$\kms$ at the turn-off, compared to 3.88~$\kms$ in model M2.
In spite of these different surface rotation rates, 
the profiles of $\Omega$ and $\theta$ are quite similar in the radiative zone of these models during the RGB
phase (Fig.~\ref{momspec}). Consequently, the turbulent diffusion
coefficients (see Eq.~\ref{eqDvTZ97}) are not very different during the RGB phase, as can be seen on
Figs.~\ref{cdiffMod2},~\ref{cdiffMod5} and ~\ref{cdiffall}.\\

This comparison shows that the structural readjustments (induced the 1st
DUP) efficiently redistribute the AM
throughout the star beyond the turn-off. The resulting rotation profile after the completion of
the 1st DUP is almost solely determined by the star's total angular
momentum at the turn-off and by the CE rotation law.

\begin{figure*}[t]
\begin{center}
\includegraphics[height=7.5cm,width=7.5cm]{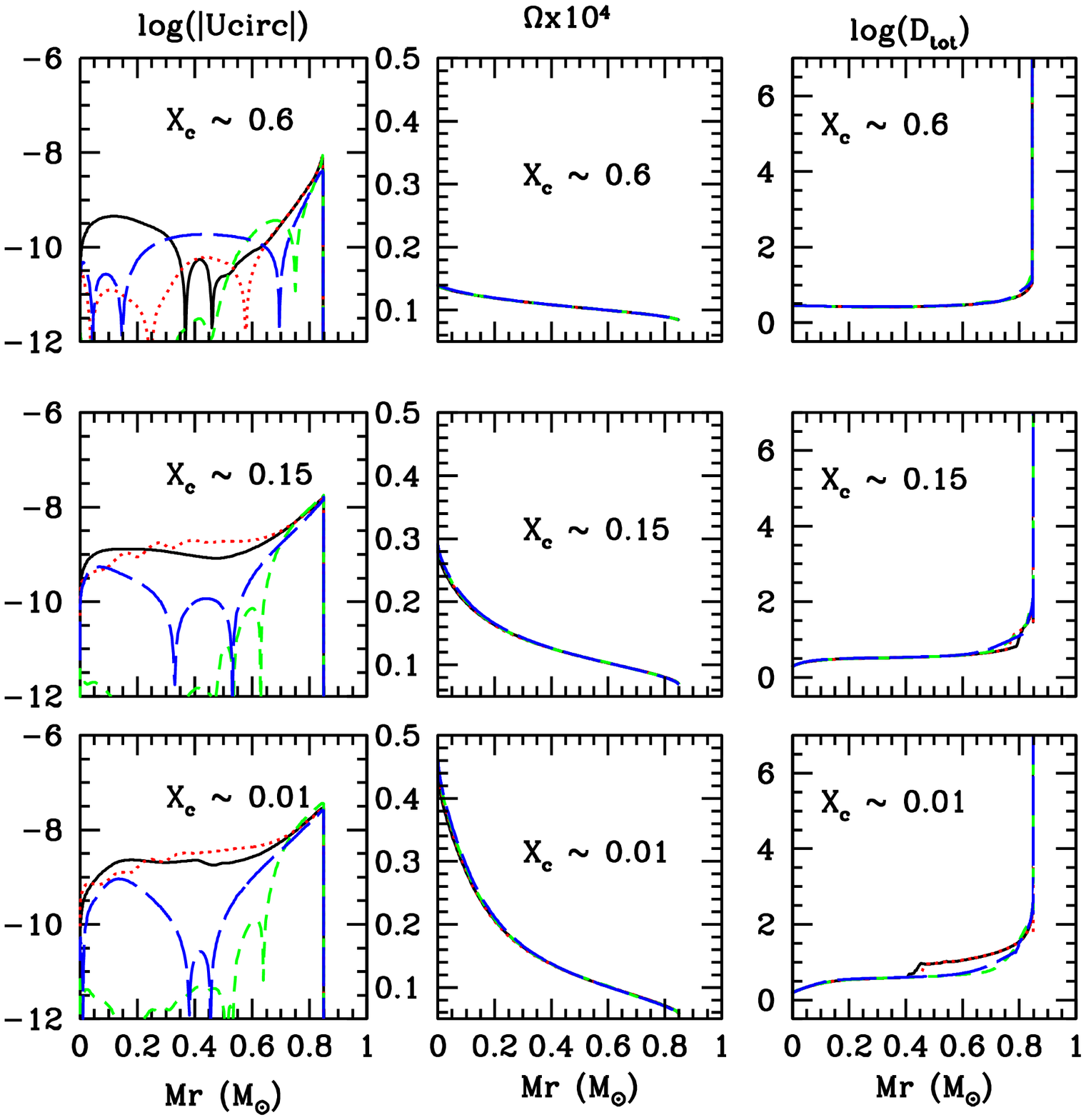}%
\includegraphics[height=7.5cm,width=7.5cm]{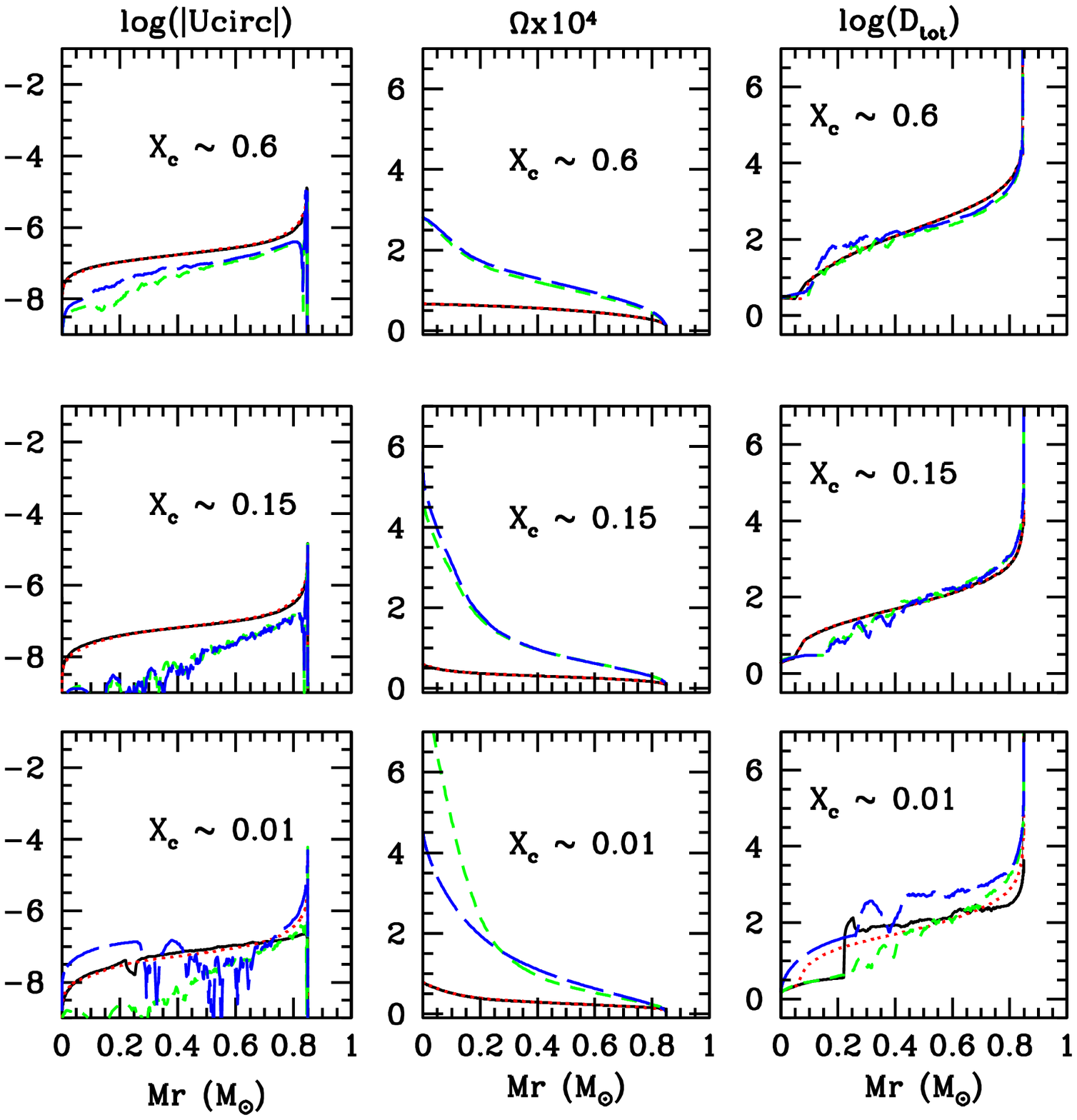}%
\end{center}
\caption{Profiles of the vertical component of the meridional
  circulation velocity $U_r$, angular velocity $\Omega$ and total
  diffusion coefficient $D_{\rm tot}$ inside $0.85 \,$\msun, [Fe/H] =
  -1.57 models at three stages of their evolution (as indicated by
  their central H content $X_c$ : $X_c = 0.6 \Leftrightarrow t \approx
  2~ Gyr$, $X_c = 0.15 \Leftrightarrow t \approx 8~Gyr$ and $X_c =
  0.01 \Leftrightarrow t \approx 10~Gyr$). {\em First 3 columns on
  the left} Profiles for models that are slow rotators on the ZAMS
  ($\upsilon_{ZAMS} = 5 \kms$) and undergo no braking.{\em Last 3
  columns on the right} Profiles of initially fast rotating models on the
  ZAMS ($\upsilon_{ZAMS} = 110 \kms$) and undergoing strong magnetic
  braking during their early MS evolution. Here, solid and dashed
  lines are superimposed and indistinguishable from one
  another. Details on input parameters are given in Table~2.}
\label{PTCF03bis}
\end{figure*}

\subsection{$\mu$-currents and horizontal turbulence}\label{subsec:mu}

The mean molecular weight affects the transport of chemicals 
via its gradient $\nabla_\mu$ and its relative horizontal variation
$\Lambda$ (see Eqs.~\ref{vcirc} and~\ref{emu}). $\mu$-gradients inhibit the
efficiency of meridional circulation, and $\mu$-currents are inhibited by
strong horizontal turbulence.
In paper~I we emphasised the importance of these terms in establishing the differential rotation profile
during the MS for Pop I stars.

In order to better assess the role of $\mu$-currents in Pop II stars,
and to determine their sensitivity to the horizontal turbulence
description, we have computed a series of models for which we alternately
use the $Zahn92$ and the $MPZ04$ prescriptions for $D_h$, assuming both
$E_\mu = 0$ and $E_\mu \neq 0$, and $\upsilon_{ZAMS} = 5 \kms$ and
$\upsilon_{ZAMS} = 110 \kms$ for each case.  The properties of these models
are summarised in Table~\ref{table3}.

\subsubsection*{Main sequence}
Let us first comment on the MS evolution. The profiles of $\Omega$, $U_r$
and $D_{\rm tot}$ are displayed on Figure~\ref{PTCF03bis} at three
different times on the MS (see Table~\ref{table3} for detailed description
of each curve).

\begin{table}
\caption{Input parameters for the models discussed in
  \S\ref{subsec:mu}. Some of the models appearing in this table are also
  listed in Tab.~1. Models Ma, Mb, Mc, Md and Mh have only been computed up
  to the turn-off. In all models,
  solid-body rotation in the CE has been assumed, and $D_v$ is given by {\em TZ97}.}
 \begin{tabular}{c | c | c | c | c | c }
 \hline
 \hline
 {\scriptsize Model} & {\scriptsize $\upsilon_{\rm
 ZAMS}$} & {\scriptsize Braking} & {\scriptsize $D_h$} & {\scriptsize
 $\mu$-currents} & {\scriptsize line style}\\
 & {\scriptsize $\kms$} & & & & {\scriptsize in Fig.~7}\\\hline
 Ma & 110  & {\scriptsize yes} &{\scriptsize  MPZ04} & {\scriptsize no} & {\scriptsize solid}\\
 Mb & 110  & {\scriptsize yes} &{\scriptsize  MPZ04} & {\scriptsize yes} & {\scriptsize dotted}\\ Mc & 110  & {\scriptsize yes} &{\scriptsize  Zahn92} & {\scriptsize yes} & {\scriptsize short-dashed}\\
 Md & 110  & {\scriptsize yes} &{\scriptsize  Zahn92} & {\scriptsize no} & {\scriptsize long-dashed}\\\hline
 M3 & 5  & {\scriptsize no} & {\scriptsize MPZ04} & no & {\scriptsize solid} \\
 M2 & 5  & {\scriptsize no} & {\scriptsize MPZ04} & yes & {\scriptsize dotted} \\
 M5 & 5 & {\scriptsize no} & {\scriptsize Zahn92} & yes & {\scriptsize short-dashed} \\
 Mh & 5 & {\scriptsize no} & {\scriptsize Zahn92} & yes & {\scriptsize long-dashed}\\\hline \hline
\end{tabular}
\label{table3}
\end{table}
{\em Slow rotators on the ZAMS (Fig. 7, left panels).}\\

Their slow rotation together with the negligible structural readjustments
occurring during the MS does not favour any mechanism able to trigger steep
rotation profiles. As a result, the transport of both angular momentum and
chemicals by rotation-induced processes is inefficient. This is independent
of the choice for the $D_h$ prescription and of the introduction or not of
the $\mu$-currents. The $\Omega$-profile as well as the total diffusion
coefficient for chemicals is similar in all cases. $\mu$-gradients in the
radiative interior are small, which makes the effects of $\mu$-currents
negligible during the MS.\\

{\em Fast rotators on the ZAMS (Fig.~7, right panels).}\\

Turning now to the fast rotators on the ZAMS, the strong braking at the
beginning of the MS allows the build-up of steeper $\Omega$-gradients, and
shear-turbulence can develop across the radiative zone already during the
main sequence. This leads to efficient transport of both angular momentum
and chemicals, similarly to what is obtained in their Pop~I counterparts.
The $\mu$-gradients are larger in this case than for slow rotators, and we
can expect $\mu$-currents to have the same effects on the rotation profile
as those observed in the Pop~I stars. It is actually the case when
considering the \emph{Zahn92} prescription for $D_h$ which was used in
paper I. As shown on Fig~\ref{PTCF03bis}, when the
$\mu$-currents are taken into account ($E_\mu \neq 0$; short-dashed lines),
the degree of differential rotation reached near the end of the MS is
substantially larger than when these terms are neglected (long-dashed
lines).  The $\mu$-currents also limit the extent of the shear unstable
region at the centre, whereas when $E_\mu = 0$, turbulence is free to
develop across the entire radiative interior (third column,
Fig.~\ref{PTCF03bis}).

Using the \emph{MPZ04} for $D_h$ leads to a different conclusion. This
prescription produces much larger values for the horizontal turbulence
$D_h$ than the former one. As the $\mu$-currents ($E_\mu$ term in
Eq.~\ref{vcirc}; see also Eq.~\ref{emu} in Appendix) are generated by
the horizontal variation of the mean molecular weight, they are much
reduced by this choice. In the case
we are considering, they are reduced to an insignificant level,
i.e. they remain small compared to the $\Omega$-currents (term $E_\Omega$ in
Eq.~\ref{vcirc}; see also Eq.~\ref{eom}) and do not affect the
rotation profile nor the meridional circulation velocity.

In short, on the MS, the $\mu$-currents have negligible effects on
Pop~II low-mass stars that are slow rotators, independently of the
prescription used for $D_h$. This is exclusively due to the slow
rotation. On the other hand, the $\mu$-currents may affect the building of
the rotation profile during the MS evolution of low-mass Pop~II stars
undergoing strong magnetic braking if the horizontal turbulence is not too
large to prevent any significant variation of the mean molecular weight to
develop.

\subsubsection*{Red Giant Branch}

\begin{figure}
\resizebox{\hsize}{!}{\includegraphics{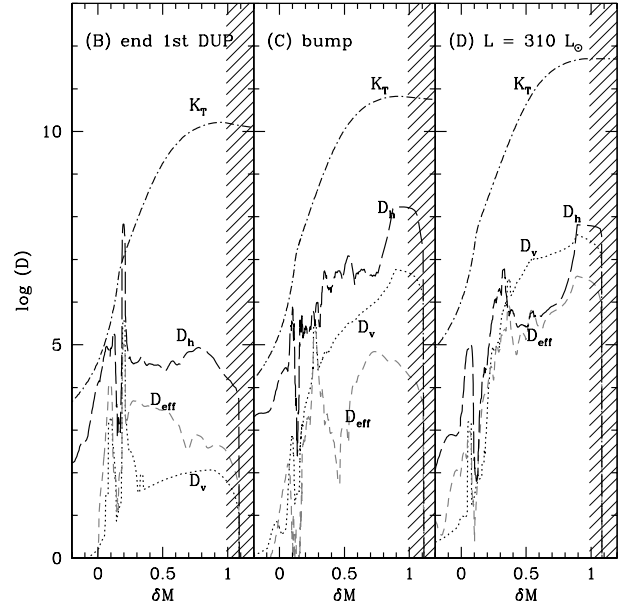}}
\caption{Same as Fig.~\ref{cdiffmodref} for M5. This
  model is similar to M2 but was computed using the Zahn (1992)
  expression for $D_h$. }
\label{cdiffMod6}
\end{figure}

Comparing model M2 with model M3 and M5 provides respective clues on the
effects of the $\mu$-currents (M2 versus M3) and the horizontal turbulence
description (M2 versus M5) on the transport of chemicals beyond the
turn-off.\\ Figure~\ref{cdiffall} presents the total diffusion coefficients
for the chemicals in all our rotating models at evolutionary points (B),
(C) and (D). $D_{\rm tot}$ in models M2, M3 and M5 (dotted, long-dashed
dotted and long-dashed lines respectively) is essentially the same at the
different evolutionary points presented. This indicates that $\mu$-currents
do not affect the transport of chemicals on the RGB. On the other hand, the
choice of the prescription for horizontal turbulence has small effects on
$D_{\rm tot}$. This is due to the fact that $D_h$ essentially affects the
effective coefficient $D_{\rm eff}$, which is much smaller than $D_v$ after
the 1st DUP in case {\em Zahn92} is used, and remains small for {\em MPZ04}.
Let us add that in general, $D_{h,MPZ04} > D_{h,Zahn92}$, and that for $D_{h,Zahn92}$, the shellular rotation
hypothesis (i.e. $D_h \gg D_v$) is violated in some regions (see panel (D)
Fig.~\ref{cdiffMod6}).

\subsection{Vertical turbulent diffusion coefficient}\label{subsec:Dv}

\begin{figure}
\begin{center}
\resizebox{\hsize}{!}{\includegraphics{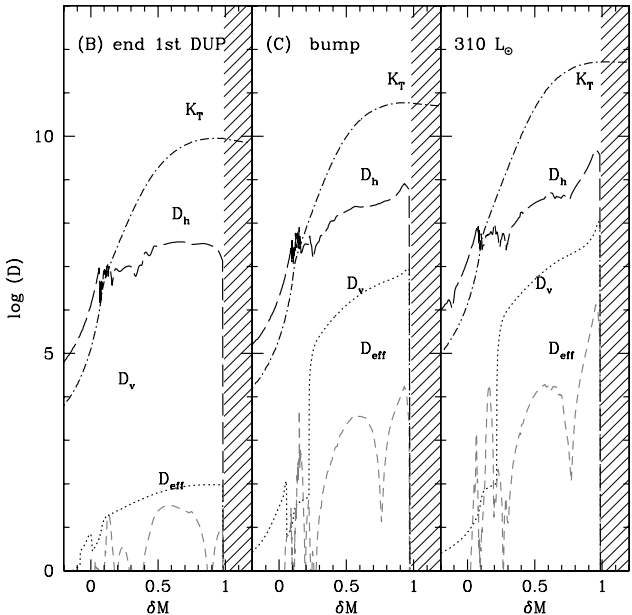}}
\end{center}
\caption{Same as Fig.~\ref{cdiffmodref} for M4. This
  model is similar to M2 except that the prescription for the vertical
  turbulent diffusion coefficient is that of Maeder \& Meynet (1996).}
\label{cdiffMod4}
\end{figure}

\begin{figure}
\begin{center}
\resizebox{\hsize}{!}{\includegraphics{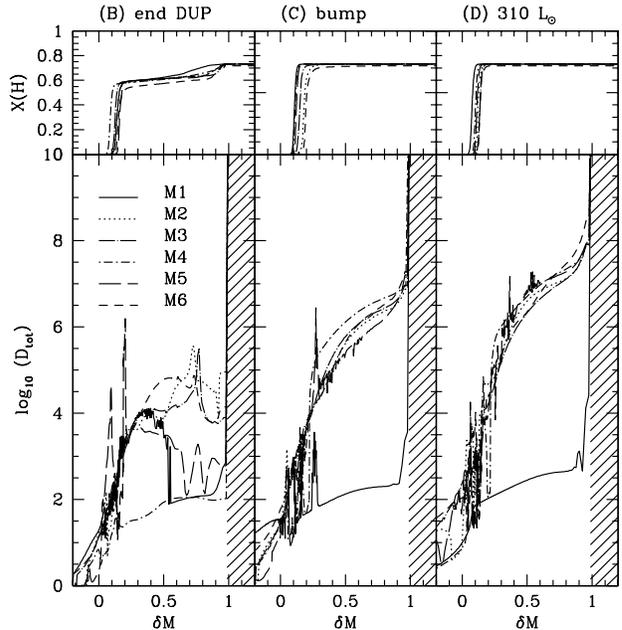}}
\end{center}
\caption{ {\em Lower panels} Total
  diffusion coefficient for several models (line encoding is given in the figure) 
as a function of the reduced mass coordinate $\delta M$ at the end of the first dredge-up and
  at the {\em bump}.{\em Upper panels} Hydrogen mass fraction profiles in
  the same models.}
\label{cdiffall}
\end{figure}

In order to compare our results with the calculations of Denissenkov \&
Tout (\cite{DT00}), we also computed a model (M4) using the \emph{MM96}
prescription for $\nu_v$ (Eq.~\ref{eqDvMM96}). The main differences between
this criterion and the one derived by Talon \& Zahn (\cite{TZ97}) were
already exposed in \S~\ref{subsec:cdiff}. In Fig.~\ref{cdiffMod4} we
present the diffusion coefficients for model M4 at the end of the 1st DUP
at the {\em bump} and at L $\simeq 310~\Lsun$. Contrary to
models M2, M3, M5 and M6, for which we also assumed uniform specific
angular momentum in the CE beyond the turn-off, by the end of the 1st DUP,
the shear instability has not developed between the base of the CE and the
edge of the HBS in model M4. Indeed Eq.~(\ref{eqDvMM96}) shows that the
shear must be larger than $N^2_\mu$ for the instability to
develop. 
In the case of giants, it is only after completion of the 1st DUP,
where the retreating CE leaves a chemically homogeneous region ($N^2_\mu = 0$), that the
shear instability may set in. However the CE remains disconnected from the
nucleosynthesis regions until the large $\mu$-barrier left by
the DUP is erased at the {\em bump}. This justifies the assumption of
Denissenkov \& Tout (\cite{DT00}), that considers rotational
transport of chemicals only from the {\em bump} on.

 However, Meynet \& Maeder (\cite{MM97}) showed that using this strong
 criterion prevents mixing from occuring in massive, fast rotating stars, in
 contradiction with observational evidence. The same conclusion that was also
 reached by Talon \etal (\cite{TZMM97}), and motivated the authors to
 introduce the erosion of the $\mu$-gradient by horizontal turbulence. For the same reason, Maeder (\cite{M97}) also
 developed a modified shear criterion to reduce the efficiency of mean
 molecular weight gradients.
 
  Being in the same framework (i.e. study of
 transport associated with meridional circulation and shear turbulence),
 and in absence of strong observational evidence, it is not justified to
 change these prescriptions for the particular case of low-mass RGB
 stars. It was SM79 who first suggested that prior to the {\em bump},
 mixing should be hindered by the $\mu$-barrier left at the end of the 1st
 DUP, and that surface abundance variations should not be expected before
 this evolutionary point. We will show however in the following that a mixing
 process eroding the $\mu$-gradient does not necessarily alter the surface abundance
pattern prior to the {\em bump}.

\subsection{Dominant process for the transport of angular momentum and chemicals}

In the framework of this paper, we consider two transport
  processes, namely meridional circulation and turbulence induced by the secular
  shear instability.
  Whereas shear-induced turbulence is always described as a diffusive
  process, meridional circulation appears as an advective process for AM and as a diffusive process for
  the chemical species (Eq.~\ref{momevol} and Eq.~\ref{diffelts}
  respectively). The relative importance of these processes can thus
  be different depending on whether we consider AM 
  or elements transport.\\
In the case of chemical species, Figs.~\ref{cdiffmodref}--\ref{cdiffMod5} and
\ref{cdiffMod6}--\ref{cdiffMod4} indicate clearly that shear-induced
turbulence is the dominant transport process beyond the completion of the
1st DUP when uniform specific angular momentum is assumed in the
envelope. The exception is model M5, where due to the smaller
efficiency of the horizontal shear turbulence ($D_h$ = $D_{h,Zahn92}$),
meridional circulation is large and still dominates the transport of
chemicals at the end of the 1st DUP (Fig.~\ref{cdiffMod6}). 
Higher on the RGB however, turbulence recovers the upper hand in this model too.
For model M1, shear-induced turbulence can not be triggered in
the radiative zone after the 1st DUP, and meridional circulation
accounts for transport for chemicals. The value of $D_{\rm eff}$ remains
however always smaller that the molecular viscosity $\nu_{mol}$.

In order to estimate the efficiency of the two processes responsible for
the AM transport, we compare their relative characteristic timescales.
The characteristic timescales for AM transport by meridional circulation and
by shear-induced turbulence over a distance $\Delta r$ can be estimated as
$\tau_U \sim
\frac{\Delta r}{U_r}$ and $\tau_\nu \sim \frac{(\Delta r)^2 \Omega}{D_v \Delta \Omega}$
respectively.\\
$\tau_U << \tau_\nu$ at the end of the 1st DUP for all models but model M1. This situation is subsequently
modified as shells with a large angular momentum are incorporated from
the inner CE into the radiative interior. From the {\em bump} on, the
efficiency of turbulence becomes thus similar to that of meridional circulation,
with $\tau_\nu \simeq \tau_U \sim 10^5$ yr. For model M1, $\tau_\nu \simeq
\tau_U \simeq 10^7$ yr from $\delta M \sim$ 0.1 to 0.8 at all times
(evolutionary points (B), (C) and (D)). In the region just
below the CE, for $\delta M \in [0.8; 1]$, the degree of differential rotation is very small (see
Fig.~\ref{omegatheta}) so that the timescale for AM transport by
turbulence is of the order of 10 Gyr, and meridional circulation
dominates.\\
 
Between the turn-off
and the completion of the 1st DUP, meridional circulation dominates the
transport of AM in all our models, while the chemical species are mainly
transported through turbulence. Beyond the 1st DUP, chemicals are transported via shear-induced turbulence while the AM
essentially evolves due to structural readjustments. Indeed at this phase,
the advective (meridional circulation) and the diffusive (shear-induced
turbulence) processes almost compensate each other, letting the Lagrangian
term control the evolution of the angular velocity profile. 
In those models where turbulence can not develop, meridional circulation dominates the
transport of chemicals and AM (model M1).\\ The dominant transport process for AM can thus differ from
the process controlling the transport of chemicals. It can also vary along
the evolution.

\section{Signatures of mixing}\label{sec:signmix}

\subsection{Structure evolution}\label{subsec:structure}

Through the transport of chemicals, and in particular those contributing to
the nuclear energy production and the opacity, rotation may indirectly affect the
evolution of stars in terms of lifetimes, luminosities and effective
temperatures. Centrifugal forces can also affect the structure but this
effect is negligible for the low rotation rates considered here.

\subsubsection{Main Sequence}

Figure~\ref{HRD} presents the Hertzsprung-Russell (HR) diagram for the models
listed in Table~\ref{table1}. Models~M0 to M3 can hardly be distinguished
on the figure. Due to their small initial rotation velocity and the
absence of braking to pump AM, models M1 to M3 present very
weak differential rotation resulting in inefficient mixing during the MS
phase 
($D_{\rm tot} \approx 10^2 - 10^3~ {\rm cm}^2{\rm s}^{-1}$); their
evolution as well as their chemical structure is thus only scarcely
affected by rotation-induced mixing.
Table~\ref{table2} presents the main evolutionary characteristics of
our models. The luminosity and effective temperature at the turn-off,
as well as the time spent on the MS are similar in slowly rotating (M1-M5)
and standard (M0) models.

 Model~M6 deviates from the standard and
slow-rotating tracks. In this model, the strong braking applied during
the first million years spent on the MS creates a large differential
rotation inside the star leading to an efficient 
transport of the chemicals
(see Eq.~\ref{eqDvTZ97}). 
In this case, as helium diffuses outwards, the opacity is globally lower
and the star consequently bluer and more luminous. As a result of fuel replenishment due to efficient rotational
mixing, the H burning phase also lasts for $\approx 400$ Myr longer in
model M6, which is thus older at the turn-off compared to
the other rotating models (Table~\ref{table2}).

 For all our rotating models, the surface velocity at the end of the MS is
lower than 6~$\kms$,
in fair agreement with the upper limits derived for globular
clusters MS stars (Lucatello \& Gratton \cite{LG03}). During the MS
evolution, the surface rotation velocity remains almost constant in models
M1 to M5. They are slowed from 5~$\kms$ on the ZAMS to $\approx 4~\kms$ at the
turn-off mainly due to the structural readjustment that become important at
the end of this phase. Model M6 undergoes magnetic braking on the MS
so that its rotational velocity 
has already dropped below $\approx 6~\kms$ when the model reaches the middle of
the main sequence (i.e. for $X_c \simeq 0.5$).
This velocity further decreases due to the
efficient transport of AM in the radiative interior, and reaches 5.7$\kms$
at the turn-off.

\begin{table}[t]
\caption{Main evolutionary features of the models presented in
  Table~\ref{table1}. TO index stands for {\it turn-off values}. 
  $M_{\rm BCE,DUP}$ is the mass coordinate associated with the deepest
  extension of the convective envelope during the first
  dredge-up. $\Delta t_{\rm bump}$ is the time needed for the entire HBS to
  pass through the $\mu$-discontinuity left by the 1st DUP.}
\begin{tabular}{c | c  c  c  c  c  c  c}
\hline \hline Model & 0 & 1 & 2 & 3 & 4 & 5 & 6 \\\hline 
$t_{\rm TO}$ {\scriptsize (Gyr)} & {\scriptsize 11.11} & {\scriptsize
11.20} & {\scriptsize 11.21} & {\scriptsize 11.21} & {\scriptsize 11.21} &
{\scriptsize 11.21} & {\scriptsize 11.61}\\\hline 
$L_{\rm TO}$ {\scriptsize ($\Lsun$)} &
{\scriptsize 3.35} & {\scriptsize 3.34} & {\scriptsize 3.35} & {\scriptsize
3.35} & {\scriptsize 3.35} &{\scriptsize 3.35} &{\scriptsize 3.55} \\\hline 
$T_{\rm eff,TO}$ {\scriptsize (K)} & {\scriptsize 6534} & {\scriptsize 6547} &
{\scriptsize 6543} & {\scriptsize 6543}& {\scriptsize 6543} & {\scriptsize
6543} &{\scriptsize 6582} \\\hline 
$\upsilon_{\rm TO}$ {\scriptsize ($\kms$)}& {\scriptsize 0} &
{\scriptsize 3.88} & {\scriptsize 3.90} & {\scriptsize 3.89} & {\scriptsize
  3.85} & {\scriptsize 3.95} &{\scriptsize 5.71} \\\hline 
{\scriptsize $M_{\rm
BCE,DUP}$ ($\Msun$)} & {\scriptsize 0.306} & {\scriptsize 0.303} &
{\scriptsize 0.304} & {\scriptsize 0.306} & {\scriptsize 0.304}  & {\scriptsize
  0.307} & {\scriptsize 0.306}\\\hline  
$L_{\rm bump}$ {\scriptsize($\Lsun$)} & {\scriptsize 107} &
{\scriptsize 101} & {\scriptsize 109} &{\scriptsize 113} & {\scriptsize 103} &
{\scriptsize 110} &{\scriptsize
  114}  \\\hline 
$T_{\rm bump}$ {\scriptsize (K)} & {\scriptsize 4899} & {\scriptsize 4919}
& {\scriptsize 4897} & {\scriptsize 4887} & {\scriptsize 4910} &
{\scriptsize 4896} &{\scriptsize 4897} \\\hline 
$\Delta t_{\rm bump}$ {\scriptsize (Myr)} & {\scriptsize 4.87} &
{\scriptsize 7.14} & {\scriptsize 7.75} & {\scriptsize 6.28} &{\scriptsize 5.75} &
{\scriptsize 6.68} & {\scriptsize 7.73} \\
\hline
\end{tabular}
\label{table2}
\end{table}

\subsubsection{Red Giant Branch}

The large increase in radius accompanying the deepening of the
convective envelope during the 1st DUP, combined with global conservation of
AM, leads to very efficient braking of the surface layers and 
spin-up of the core. All our rotating models have surface velocities lower than
1$\kms$ at the end of the dredge-up (including model M6 for which we
  have $\upsilon_{ZAMS} = 110 \kms$). In models M1, M2, M4 and
M6, for which the total diffusion coefficient for chemicals is larger than the molecular viscosity
at the base of the envelope, the 1st DUP is 
deeper compared to the standard case. These variations of the depth of the
1st DUP (Tab.~\ref{table2}) remain however small and
will not affect significantly the surface abundance patterns at this phase
(see \S~\ref{subsec:abund}, Fig.~\ref{obsvsmodels}).

\begin{figure}[t]
\resizebox{\hsize}{!}{\includegraphics{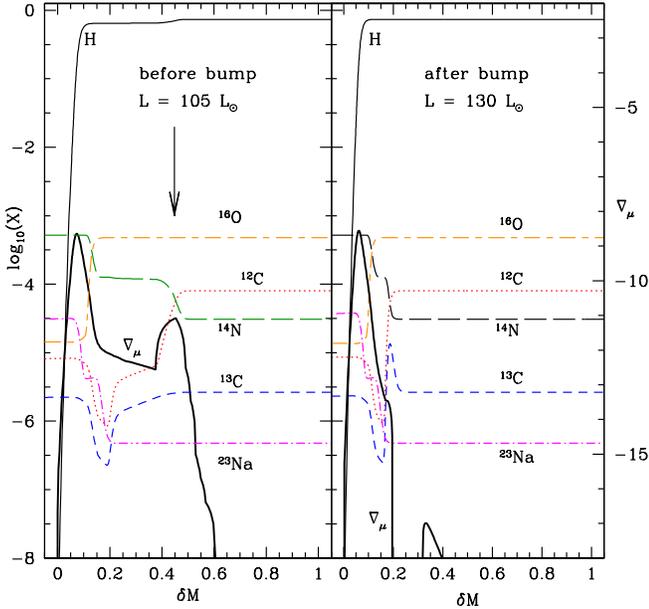}}
\caption{Logarithm of the mass fraction profiles of H, $^{12}{\rm C}$,
  $^{13}{\rm C}$, $^{14}{\rm N}$, $^{16}{\rm O}$, $^{23}{\rm Na}$ and of
  the mean molecular weight gradient $\nabla_\mu = d \ln \mu/ dr$ (bold
  line) as a function of the reduced mass inside a $0.85 \,$\msun, $Z = 0.0005$
  standard stellar model M0. The panels represent the profiles of these
  elements in the HBS just before (i.e. before the HBS contacts the
  $\mu$-discontinuity left by the 1st DUP) and after (i.e. once the mass
  coordinate at the base of the HBS is larger than $M_{\rm
BCE,DUP}$) the {\em bump}. In the
  left panel, the down-going arrow indicates the mass coordinate of maximum
  penetration of the CE during the 1st DUP.}
\label{profbumpmod0}
\end{figure}

\begin{figure}[t]
\resizebox{\hsize}{!}{\includegraphics{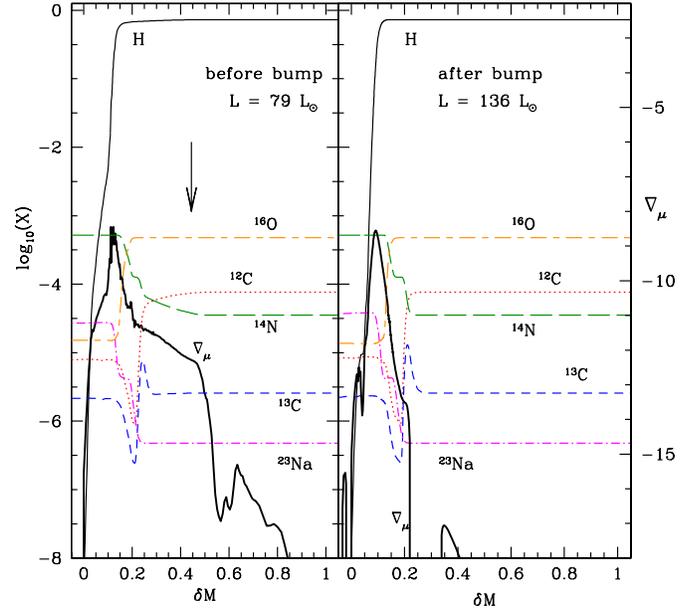}}
\caption{Same as Fig.~\ref{profbumpmod0} for our rotating reference model
  M1.}
\label{profbumpmod1}
\end{figure}

In the following we will refer to the \emph{bump luminosity} as the
\emph{luminosity of the model when the mass coordinate of the maximum
energy production inside the HBS is equal to ${\rm M}_{\rm BCE,DUP}$}. 

\begin{figure}[t]
\resizebox{\hsize}{!}{\includegraphics{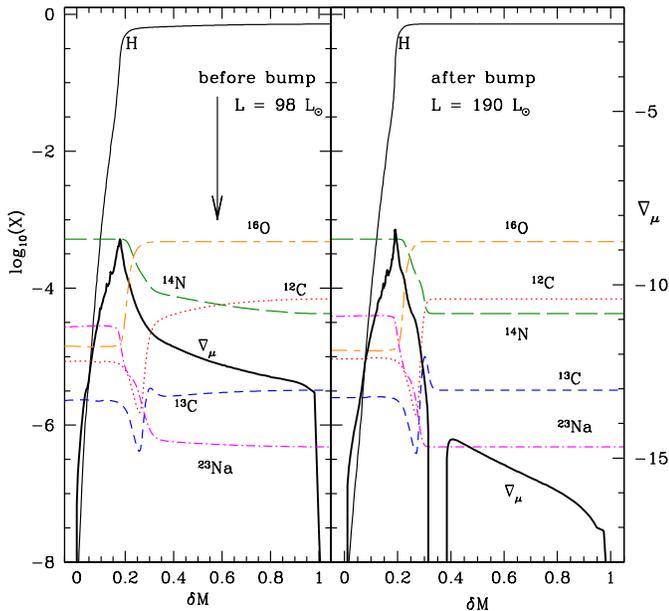}}
\caption{Same as Fig.~\ref{profbumpmod0} for model M6.}
\label{profbumpmod5}
\end{figure}

Figures~\ref{profbumpmod0}, \ref{profbumpmod1} and \ref{profbumpmod5} present
selected abundance profiles and the mean molecular weight gradient
$\nabla_\mu = \partial \ln \mu / \partial r $, before and after the {\em
bump} inside models M0, M1 and M6 respectively.  
The peak amplitude $\nabla_{\mu,c}$
in the $\mu$-gradient profile at $M_r = M_{\rm BCE,DUP}$ (indicated by
an arrow on the left panels) decreases with increasing degree of
mixing.
In the non-rotating model, this peak is a signature of the deepest penetration
of the convective envelope during the first dredge-up, and the corresponding 
$\nabla_{\mu,c}$ value is $\simeq 10^{-11}$ (Fig.~\ref{profbumpmod0}). 
In the rotating models, the peak is spread out due to the ongoing transport 
of the chemicals and varies between $\nabla_{\mu,c} \simeq 3\cdot 10^{-12}$ 
(Fig.~\ref{profbumpmod1}) and 
$\nabla_{\mu,c} \simeq 5\cdot 10^{-13}$ (Fig.~\ref{profbumpmod5}). The
amplitude of $\nabla_{\mu,c}$ directly reflects the strength
of the diffusion coefficient in this area (see
Figs.~\ref{cdiffmodref} and \ref{cdiffMod5}).  
According to Charbonnel et al. (\cite{CBW98}) the
regions where $\nabla_\mu \geq 1.5 \times 10^{-13}$ should not be 
affected by the extra-mixing acting below the CE. Therefore, even the low value of
$\nabla_{\mu,c}$ found in model M6 can prevent mixing to act freely below $\delta M \approx 0.5$. \\
Let us finally emphasise that
although the amplitude of the diffusion coefficients may be locally
increased due to numerical instabilities, and artificially lower the mean
molecular gradient in that region, the
reproducibility and constancy of the $\mu$-barrier spread over indicates
that this feature is not a numerical artifact.

\begin{figure*}[t]
\begin{center}
\includegraphics[angle=-90,width=15cm]{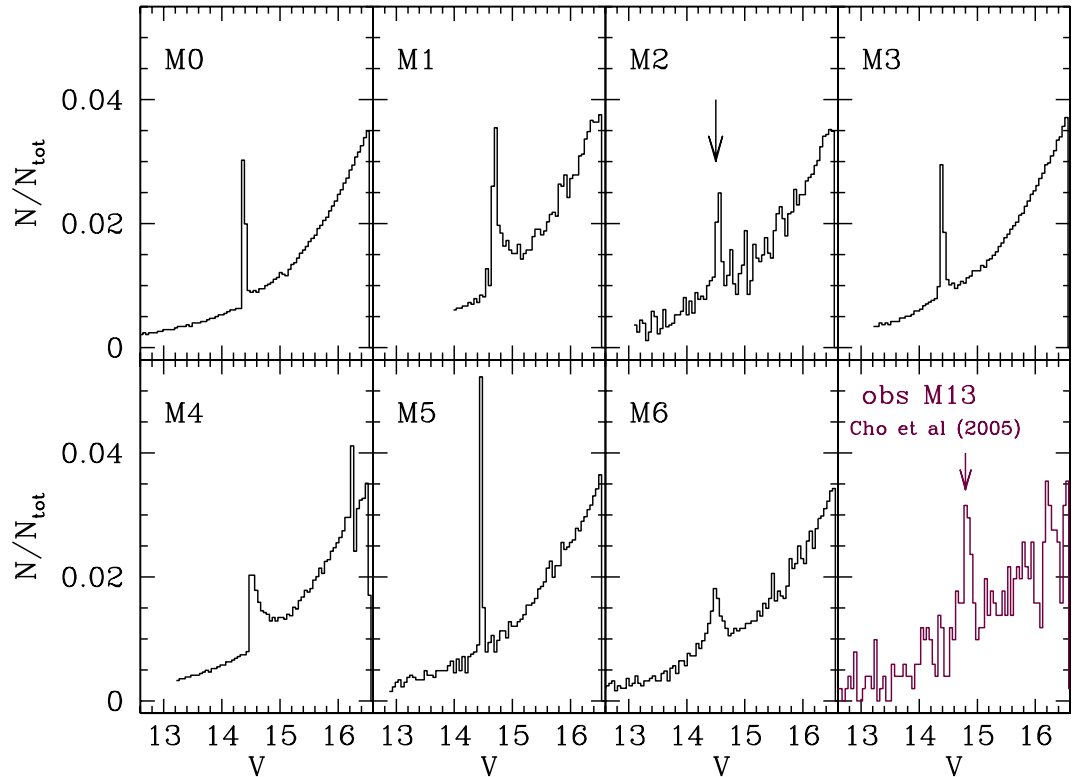}
\caption{Normalised theoretical luminosity function in V for the models presented in
  Table~1 and 3. These histograms represent the
  time spent in each magnitude (luminosity) bin as a function of
  visual magnitude V. The ordinates are normalised to the total time spent
  on the portion of the RGB where V $\leq 16.6$. This limit has been chosen in
  order to have the same normalisation as in the observational case
  presented in the $8^{th}$ panel (low right). The observed normalised
  luminosity function is based on Cho et al. (2005). The arrows in panels 3
  and 8 indicate the position of the {\em bump}.}
\label{lumfunctheo}
\end{center}
\end{figure*}

In order to evaluate the ``observational'' impact of rotational mixing
  on the luminosity function's {\em bump}, we have
  computed theoretical luminosity functions (hereafter LF) for each of our models
  (see Fig.~\ref{lumfunctheo}). These theoretical LF represent the time spent in each bin of magnitude V
  by  models on the RGB with magnitudes lower than $V =
  16.6$. We have chosen this cut-off in order to be able to compare the
  models predictions with the observed luminosity function of the
  globular cluster M13 as given by \cite{Cho05}.
  
In order to determine the $V$ magnitude associated with a given luminosity $L$, we 
used the classical magnitude-luminosity relation $$V = M_{bol,\odot}
  - 2.5 \log(L/L_{\odot}) - {\rm BC} + (m-M)_V.$$  For M13, we adopt a distance modulus
$(m-M)_V = 14.48$ , following \cite{Cho05}, and a bolometric correction ${\rm BC} =
-0.253$, according to the value given by \cite{Girardi02} for [Fe/H] $=
-1.5$, $T_{\rm eff} = 5000~{\rm K}$ and $\log g = 2$. $M_{bol,\odot} =
4.76$ according to Cox (\cite{Cox2000}).

Figure~\ref{lumfunctheo} emphasises the following points :
\begin{enumerate}
\item The {\em bump} clearly appears for all models in spite of the
  numerical noise present at higher magnitudes in models M2
  and M6. When the $\mu-$barrier has been efficiently eroded, the bump is
  less pronounced in the LF.
\item The predicted LF for models with low degree of mixing (namely M1, our
  reference model, and M3, with $E_\mu = 0$) are similar to the one
  obtained for the standard model M0. In case of model M3, the
  diffusion coefficient in the outer HBS remains always small prior to the
  {\em bump} (i.e. $D_{\rm tot}\leq 5.10^3 ~{\rm cm^2.s^{-1}}$). This
  allows preservation of a large $\mu$-gradient which is responsible for
  the clear signature in the LF.  A similar effect explains the LF for
  model M1, where transport efficiency is always low in the radiative zone.
\item In rotating models, the {\em bump} occur at higher
luminosity compared to the standard model M0.

\end{enumerate}

\subsection{Abundances}\label{subsec:abund}

In \S~\ref{subsec:structure}, we mentioned the lowering of the mean
molecular weight barrier left by the 1st DUP in rotating models M1 and
M6. Figure~\ref{profbumpmod1} shows the erosion of
$^{12}{\rm C}$ and $^{14}{\rm N}$ profiles as a result of rotational
mixing. However, this chemical diffusion does not affect significantly the surface
  abundances, and we report only a minor increase of
$^{14}{\rm N}$ anti-correlated with a decrease of $^{12}{\rm C}$ compared
to the standard case.

In Model M6 the higher degree of differential rotation at the base of the
CE feeds the turbulent shear-induced mixing. The outer $^{14}{\rm N}$
plateau is erased and the nitrogen mass fraction in the CE is increased
relative to the standard case. $^{13}{\rm C}$ also diffuses outwards and
the peak around $\delta M = 0.3$ is flattened out. Although the mean
molecular weight at the depth of deepest penetration of the CE is much
lower than in the standard model, $D_v$ (which dominates the transport)
decreases rapidly from $10^9\, {\rm cm}^2{\rm s}^{-1}$ just below the CE
down to $\approx 10^5\,{\rm cm}^2{\rm s}^{-1}$ in the chemically
inhomogeneous regions of the outer HBS (Fig.~\ref{cdiffMod6}).  Mixing is
thus confined to a narrow region located just below the CE, where the
chemical profiles are flat. According to Charbonnel (1995) and Denissenkov
\& VandenBerg (\cite{DV03}), diffusion coefficients as large as $\approx 4
\cdot 10^8 \,{\rm cm}^2{\rm s}^{-1}$ are needed to connect the HBS with the
CE and to modify the surface abundances. However this configuration is never
reached in our self-consistent models.

\begin{figure*}[t]
\begin{center}
\includegraphics[width=12cm]{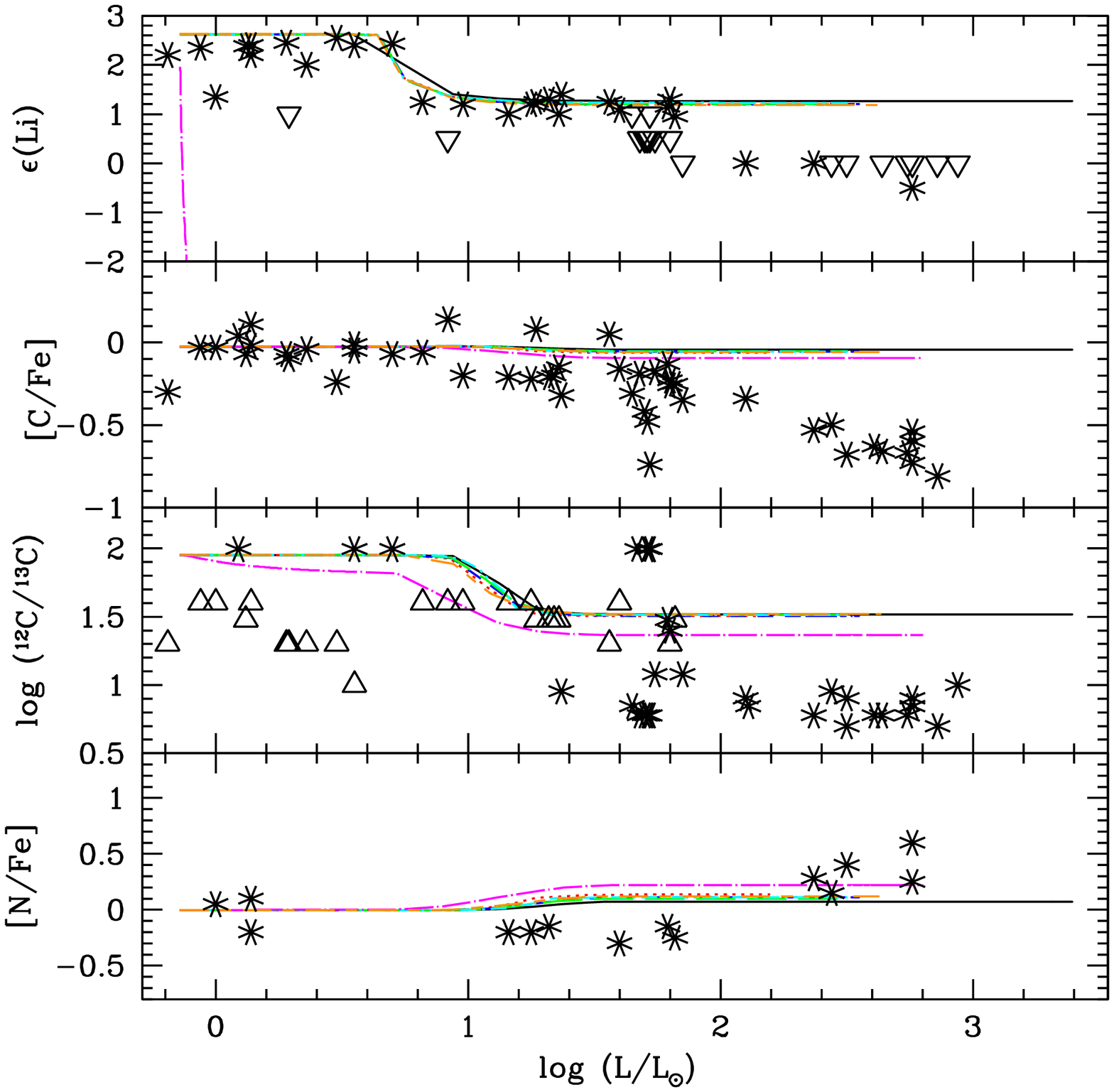}
\caption{ Evolution of the lithium abundance ($\epsilon
  (^{\rm 7Li})$), the carbon isotopic ratio \cratio, [C/Fe] and [N/Fe] as a
  function of the luminosity logarithm for the models presented in
  Table~1. Line encoding is the same as in Fig.~\ref{HRD}. Black asterisks are actual 
  measurements, open upward triangles are {\em lower}
  limits and open downward triangles are {\em upper} limits. Observational data from Gratton et
  al. (2000) for field stars in the metallicity range $[{\rm Fe/H}] \in
  [-2;-1]$.}
\end{center}
\label{obsvsmodels}
\end{figure*}

We reach similar conclusions for the other rotating models, since at the {\em bump}, the
diffusion coefficients have the same magnitude,
and are too small to affect the surface abundance composition. This is clearly
illustrated in Fig.~\ref{obsvsmodels}, where we present a comparison of the
temporal evolution of lithium, carbon, nitrogen and carbon isotopic ratios
obtained in our models with homogeneous observational data for field stars
with [Fe/H] $\in [-2; -1]$. While variations associated with the 1st DUP are
satisfactory and rotating models tend to better agree with observations, no further
variations are obtained after the {\em bump}. Let us note that model M6 leads to
the destruction of lithium already on the MS (because of strong mixing associated with
the shear in absence of any compensating mechanism) in contradiction with the
observations. 
  This situation is similar to that encountered in Pop~I stars
on the red side of the lithium dip. For a complete discussion of this problem,
the reader is referred to Talon \& Charbonnel (\cite{TC98}), as well as to Talon \& Charbonnel (\cite{TC05}) and Charbonnel \& Talon (\cite{CT05}) which describe how internal gravity waves could help resolve this issue in Pop~II stars.

 Regarding lithium on the RGB the mixing rates associated with
shear-induced turbulence do not allow the triggering of the Li-flash as
proposed by Palacios et al. (\cite{PCF01}) to consistently explain the low
percentage of lithium rich giants at the {\em bump} luminosity. Indeed, in
this scenario, an initial mixing rate of about $10^9 {\rm cm}^2.{\rm
s}^{-1}$ is needed in the region of the $^7{\rm Be}$ peak region in order
for it to diffuse outwards and decay into $^7{\rm Li}$ in a region where
this nuclide will efficiently capture protons and give rise to a energetic
runaway called the ``Li-flash''. We never get such high mixing rates in the
radiative interior of our models.\\ The present description of the
extra-mixing process in RGB stars does not allow the models to reproduce the observed
abundance anomalies in upper RGB stars, and does not validate the Li-flash
scenario for Li-rich RGB.

\section{Comparison with previous works}\label{sec:sumprevwork}

In this paper we have presented the first models of rotating
low-mass stars which take into account rotational transport
by meridional circulation and shear turbulence coupled self-consistently to the
structural evolution from the ZAMS to the upper RGB.
A detailed study of the input physics associated with the rotational
transport of angular momentum and chemicals allowed us to assess the impact
of various physical ingredients on the extension and magnitude of mixing
along the giant branch. 
Let us know compare our predictions with others from the literature
\\

\begin{enumerate}

\item {\underline {\bf Angular momentum evolution}}\\

With regard to the evolution of the angular velocity profile, we assumed
solid body rotation on the ZAMS in all our models and then let angular
momentum be transported by meridional circulation and shear-induced
turbulence. At the turn-off, this leads, in all cases, to differential
rotation in the radiative interior and to a slowly rotating convective
envelope. {\em Beyond the turn-off, the angular velocity below the CE is
determined by the assumed rotation law in the CE. Both the absolute value
of $\Omega_{\rm BCE}$ and the degree of differential rotation (and hence
mixing) in the radiative zone are maximised after the completion of the 1st
DUP in the case of a differentially rotating CE ($j_{\rm CE}(r) =
cst$)}. This supports the conclusions anticipated by other authors. In
1979, Sweigart \& Mengel conjectured that differential rotation in the
convective envelope of a red giant with homogeneous specific angular
momentum could be necessary to provide enough rotational mixing at the {\em
bump}. Recently, Chanam\'e \etal (\cite{CPT05}), using a ``maximum mixing
approach'', reach the same conclusion from considerations on the global AM
budget. They also indicate that if $\Omega$ has to be described by a power
law, it is not necessarily with a -2 index (e.g. uniform specific angular
momentum). As we mentioned in \S~\ref{subsec:rotCE}, assuming uniform
specific angular momentum is a first approximation, and we may expect
better estimates of the CE rotation regime from direct numerical
simulations.\\ 
In their work, Chanam\'e \etal (\cite{CPT05}) nonetheless insist on the fact that in
order for rotational mixing to reproduce the observed abundance anomalies
of low-mass Pop II RGB stars, they need to assume unrealistic large
rotation velocities at the turn-off. This is a conclusion that we also
reach in our complete computations. Differential rotation in the radiative
region separating the HBS from the CE increases as the CE retreats after
the DUP. However, all together, {\em the self-consistent evolution of the
rotation profile for realistic surface velocities at all phases leads to
transport coefficients too low by 3 orders of magnitude compared to what is
expected from parametric studies (Weiss \etal \cite{WDC00}, Denissenkov \&
VandenBerg \cite{DV03}) in order to alter the surface chemical composition
beyond the {\em bump luminosity} and reproduce the observed patterns.}

In this work, we did not force the specific angular
  momentum to {\em have the same value at all times in the convective envelope}
  , as was done by Denissenkov \& Vandenberg (\cite{DV03}). Under such an
  assumption the specific angular momentum within the
  CE remains at the same level during the 1st DUP as at the turn-off, which results in an
  increase of the differential rotation and mixing below the CE. On the
  contrary, in the models M2 to M6 presented here, when the CE deepens
  during the 1st DUP, it dredges material with lower specific angular
  momentum, and $j_{CE}$ (which is the same in each mass shell within the
  CE) drops as can be seen from Fig.~\ref{momspec}. As a result, the
  angular velocity and the differential rotation are lower at the {\em bump}. Beyond the completion
  of the 1st DUP, $j_{CE}$ continues to evolve slightly due to angular
  momentum transfer with the underlying radiative zone, but does not
  significantly vary anymore. Assuming {\em no variation in space} of the
  specific angular momentum in the CE after the turn-off appears to be very different from assuming
  {\em no variation with time} of this same quantity. This latter
  assumption, when combined to strong differential rotation
  in the radiative interior on the MS, is an {\em ad hoc} way to produce
  strong differential rotation (and mixing) at the {\em bump}.

Although Denissenkov and collaborators (Denissenkov \& Tout
  \cite{DT00} ($DT00$), Denissenkov \& VandenBerg \cite{DV03} ($DV03$)) 
  have also searched for a solution to the RGB abundance anomalies problem 
  in terms of rotational mixing by meridional circulation and shear
  turbulence, they reach very different conclusions in terms of the
  evolution of the surface abundance pattern on the RGB. 
  As a matter of fact, their transport coefficients are very similar to ours in terms of shape,
  but they get at all phases much larger rotation velocities and differential rotation
  rates than we do, resulting in higher mixing rates.
  
  The origin of the differences is difficult to assess
  although several points are certainly critical.
  First of all the self-consistency of the treatment of rotational transport within the
  stellar evolution code, which allows the retro-action of AM and chemicals
  transport on the structure at each evolutionary step seems to be crucial. 
  Although $DT00$ solved Eq.~\ref{momevol} using Maeder \& Zahn (\cite{MZ98}) formalism, 
  this was done outside their evolution code in a post-processing way. 
  In addition they imposed an angular velocity profile at the
  {\em bump} (up to this point their model 
  does not consider any transport) 
  so as to reproduce the observed anomalies in the globular cluster
  M92. 
  As a result they obtain very large mixing rates able to
  change the O and Na surface abundances.  As mentioned in
  \S~\ref{sec:intro}, recent observations in different GCs indicate that
  the O-Na anti-correlation also exists in turn-off stars, with a similar
  spread as in RGB stars. This strongly suggests that this pattern is
  predominantly of primordial origin, and that there is no need for the
  evolutionary models to reproduce it, at least when considering an average
  GCRGB star\footnote{In the specific case of M13, the extreme O-Na
  anticorrelation observed in RGB tip stars could on the other hand be attributed to some
  extreme mixing whose signature superimposes to the primordial pattern and
  exacerbates it.}.
  In our approach, the
  angular velocity profile at the {\em bump} is not assumed but
  results from the evolution (due to structural readjustments
  and rotational transport) from the ZAMS. It is by no means a free
  parameter that can be tuned at the {\em bump}.

  Another important difference concerns the assumption made by $DV03$ on the evolution of the
  angular velocity at the base of the CE (in this paper they do not 
  consider the rotational transport of AM, and they apply the rotational 
  transport to the chemicals only from the {\em bump} on).
  In order to get the ``right'' $\Omega$ profile at the {\em bump}, they need to impose the constancy of 
  the specific angular momentum {\em from the ZAMS up to the RGB tip}.
  We consider that this strong assumption is unphysical.\\
  
  As a consequence, the work by Denissenkov and collaborators should not be
  considered as a proof of the efficiency of rotational transport by
  meridional circulation and shear-induced turbulence to modify the surface
  abundance pattern of low-mass RGB stars. It however shows
  what should be the angular momentum distribution at the {\em bump} in a
  rotating star, for shear-induced turbulence to produce the expected amount
  of mixing required in these objects.\\

\item \underline{\bf Description of the turbulence}\\

   We have investigated the effects of using different prescriptions for
  both the horizontal and vertical turbulent diffusion coefficients during the
  RGB evolution. {\em Although some differences arise depending on the
  adopted descriptions of turbulence, their effects on the transport of
  chemical species at the {\em bump} and beyond are marginal}. Concerning
  the choice for $\nu_h$, the {\em MPZ04} prescription, predicts a
  larger value which ensures the validity of the shellular rotation scheme,
  a condition that is not always fulfilled when using {\em Zahn92} prescription. The {\em MPZ04} prescription also
  (over-)quenches the efficiency of meridional circulation and that of the
  $\mu$-currents.  Regarding the choice for $\nu_v$, contrary to the
  expectations of Denissenkov \& Tout (\cite{DT00}), the \emph{TZ97}
  prescription for $\nu_v$ does not lead to efficient mixing throughout the
  entire radiative zone, nor does it contradict the observations when
  implemented in a self-consistent scheme where $\mu$-gradients are taken
  into account. The observations in low-mass RGB stars do not allow any
  discrimination between the \emph{MM96} and the \emph{TZ97} prescriptions
  for $\nu_v$, and the use of the former prescription in order to prevent
  any mixing in the outer HBS prior to the {\em bump}, as advocated by
  Denissenkov and collaborators, is not justified. Let us also recall that
  the \emph{TZ97} prescription gives a much better agreement in the case of
  massive stars.\\
 
\item \underline{\bf $\mu$-currents and $\mu$-gradients}\\

Several important results obtained from our models concern the effect of
mean molecular weight gradients ($\mu$-gradients) on the rotational
transport.\\ 

In this paper and in paper I, we have studied the effect of the $\mu$-gradients
on the main sequence. In \S~\ref{subsec:mu}, we have shown that if the
present low-mass RGB stars were slow rotators on the ZAMS,
shear-induced turbulence could not develop in the radiative interior during
the main sequence. In this case, and independently of the prescription used
for the turbulent diffusion coefficients, the $\mu$-currents have no
effects since rotational mixing is negligible. For model M6, a fast
rotator on the ZAMS undergoing strong braking on the main sequence, we
reach the same conclusion as in paper~I when the {\em Zahn92} prescription
is used for $D_h$: $\mu$-currents play an important role in shaping the
turn-off rotation profile. On the other hand, the use of the {\em MPZ04}
prescription for $D_h$ reverts this conclusion and the $\mu$-currents are
insignificant, as in the case of the slow rotators.

 {\em Beyond the turn-off, the transport erodes the $\mu$-gradients in all
our rotating models, including those with uniform angular velocity in the
CE.}  The $\mu$-discontinuity translates into a dip in the diffusion
coefficient profiles. The lesser the mixing, the broader and the more
persistent this feature. In models with uniform specific angular momentum
in the CE, this gap is soon filled after the 1st DUP because turbulent
transport is efficient enough in this region to smooth the chemical
gradients. Despite the lowering of the $\mu$-barrier, $D_{\rm tot}$ remains
however too small to connect the outer HBS with the CE, and the surface
abundance pattern is not altered prior to the {\em bump}. Concerning
the observational consequences, the spreading over of the
$\mu$-barrier lowers the luminosity function height at the {\em bump}, but does
not erase it. {\em The $\mu$-gradients are thus seemingly not entirely
responsible for the lack of mixing evidence in lower RGB stars, contrary to
what was conjectured by SM79 and Charbonnel (\cite{CC95}). A similar
conclusion was reached by Chanam\'e et al. (\cite{CPT05}) in their
``optimised rotational mixing'' approach.}\\

Finally, in all our rotating models, the $\mu$-barrier associated with the
HBS very efficiently prevents any mixing to connect the Na-rich layers with
the outer radiative envelope. Thus, contrary to Chanam\'e \etal
(\cite{CPT04a}, \cite{CPT04b}, \cite{CPT05}) and Denissenkov \& VandenBerg
(\cite{DV03}) the mixing depth does not need to be parametrised if the
effects of $\mu$-currents on the mixing are consistently taken into
account.

\end{enumerate}

\section{Conclusion}\label{sec:conclusion}

Our self-consistent approach of rotational mixing associated with
meridional circulation and shear-induced turbulence leads to two major
conclusions : (1) {\em this formalism does not provide large enough
transport coefficients in low-mass, low-metallicity RGB stars as required
to explain the abundance anomalies observed both in the field and in
globular clusters}; (2) {\em it requires differential rotation in the
convective envelope}\footnote{This rotation regime is achieved in our case
by imposing uniform specific angular momentum within this region.}{\em in
order to obtain non-negligible differential rotation rates, and hence
mixing rates, in the underlying radiative region}.

These results point toward remaining open questions that we would like to
bring to light.\\

The interplay between convection and rotation in extended stellar
convective envelopes is still unknown, and the hypothesis of differential
rotation is very attractive when the shear is the only process considered to trigger
the turbulent transport of angular momentum and chemicals.\\
In our work, it appears that changing the angular velocity profile in the
convective envelope from uniform to differential, increases the degree of mixing in the
underlying radiative region. This enhancement remains however moderate and
does not lead to the large diffusion coefficients expected from parametric
studies. As the shear-induced turbulence appears not to be efficient {\em
  on its own} to reproduce the observed abundance anomalies in low-mass
giants independently of the rotation law in the convective envelope, we are
not able at present to make any statement concerning this aspect. We
nonetheless would consider
with great interest the use of 3D hydrodynamical direct simulations to
assess the rotation regime within the extended convective envelopes of
cool giants.

In their recent work, Chanam\'e \etal (\cite{CPT05}) 
propose that the rotation regime of the convective envelopes may
change along the evolution, going from solid-body during the MS to
differential rotation on the RGB. 
Such a scheme would reconcile rotation
velocities of MS and horizontal branch stars together with abundance
patterns from the MS to the RGB tip. 
Here again, if another physical process, such as internal gravity waves,
is able to transport angular momentum also in giants and if its
efficiency at all evolutionary phases depends on the initial
mass, the modification of the rotation regime in the convective envelope
during the giant phase could not be necessary anymore.

The second point concerns the transport mechanisms associated
with differential rotation. By now, only the secular shear instability has been
investigated. This is mainly related to historical reasons, since Zahn's
formalism was at first derived for MS stars, where this
hydrodynamical instability is dominant. The structure of a
giant star is however very far from resembling that of the Sun. We might
then expect that specific features such as nuclear burning shell,
contracting radiative interior and expanding (extended) convective envelope
favour the triggering of other hydrodynamical instabilities, or physical transport
mechanisms. Spruit \& Knobloch (1984)
advocated the possibility for the baroclinic instability to be efficient in
giant stars, but at present, we lack a description in the non-linear
regime (e.g. in a regime associated with turbulence). This instability also
depends on the degree of differential rotation but is not as sensitive to
$\mu$-gradients as the secular shear, and could thus complement the effect
of secular shear to increase the degree of mixing in regions with large
$\mu$-gradients. Other instabilities such as the Goldreich-Schubert-Fricke
(GSF) and the Solberg-H\o iland instabilities could also become
non-negligible during the RGB phase.

Last but not least, and in connection with the presence of the 
so-called ``super Li-rich giants" at the RGB {\em bump} (Charbonnel \& Balachandran \cite{CB00} 
and references therein), Palacios et al. (\cite{PCF01}) proposed that the structural and 
nuclear response of the star to the rotation-induced mixing could cause an increase
of $D_{\rm tot}$ as required to explain the abundance anomalies. This scenario 
involves an important release of energy due to $^7$Li burning in the 
external wing of the HBS (the so-called ``Li flash"). This opens a new 
field of investigation concerning the reaction of meridional circulation
and of the various instability to a major and local release of nuclear energy.

Our results clearly show the lack of success of secular shear {\em alone} to
trigger large enough amount of extra-mixing, but does
not rule out rotation-induced mixing being responsible
for abundance anomalies in low-mass giants. As a matter of fact, rotation remains
at present the best candidate for the triggering of the extra-mixing process
in these objects. Our study points toward the
need for more physics, implying a revision of the coupling between rotation
and convection, and the treatment of other physical processes
likely to contribute to transport angular momentum, including 
hydrodynamical instabilities neglected before.

\begin{acknowledgements}
The comments of the referee allowed improvement of the initial
version of this paper. 
LS is FNRS research associate. AP acknowledges financial support from the ESA PRODEX
contract 96009. CC is supported by the Swiss National Science Foundation.
We thank the French Programme de Physique Stellaire (PNPS) and Programme
National Galaxies (PNG) for travel support.

We dedicate this work to the memory of our friend Dr. Manuel Forestini.

\end{acknowledgements}

\appendix{}
\section{Detailed expressions for the components $E_\Omega$ and $E_\mu$ in Eq.~\ref{vcirc}}

Maeder \& Zahn (\cite{MZ98}) derived a generalised expression of the
meridional circulation velocity $U_r$, taking into account the effects of
$\mu$-gradients, non-stationarity and of a general equation of
state, that was revised and corrected in Denissenkov \& Tout (\cite{DT00}).

Here, we reproduce the detailed expressions of the $E_{\Omega}$ and $E_{\mu}$
terms appearing in Eq.~\ref{vcirc}. Equations \ref{eom} and \ref{emu}
correspond to corrected versions of Eq.~4.30 and Eq.~4.42 found in Maeder \& Zahn
(\cite{MZ98}).

\begin{eqnarray}
E_{\Omega} & = & 2 \left[1-\frac{\overline{\Omega^{2}}}{2 \pi G \overline{\rho}}
-\frac{(\overline{\varepsilon}+\overline{\varepsilon}^{grav})}{\varepsilon_m}\right]\frac{\tilde{g}}{\overline{g}}\nonumber\\
& - & \frac{\rho_m}{\overline{\rho}} \left\{
\frac{r}{3}\frac{d}{dr}\left[H_T \frac{d}{dr} \lp \frac{\Theta}{\delta} \rp
-\chi_{T} \frac{\Theta}{\delta} + \lp 1-\frac{1}{\delta} \rp \Theta \right]
\right. \nonumber\\ & -& \left. \frac{2H_T}{r} \left( 1+\frac{D_h}{K}
\right) \frac{\Theta}{\delta} +\frac{2}{3}\Theta \right\} -
\frac{(\overline{\varepsilon}+\overline{\varepsilon}^{grav})}{\varepsilon_m}
\left[H_T \frac{d}{dr} \lp \frac{\Theta}{\delta}\rp \right. \nonumber\\ & +
& \left. \lp f_{\varepsilon}\varepsilon_T-\chi_T \rp \frac{\Theta}{\delta}
+ \lp 2-f_{\varepsilon}-\frac{1}{\delta}\rp \Theta -
\frac{\overline{\Omega^2}}{2 \pi G \overline{\rho}}\Theta \right] \nonumber
\\ & - & \frac{\overline{\Omega^{2}}}{2 \pi G \overline{\rho}} \Theta +
\frac{M_*}{L} \frac{C_p T}{\delta} \frac{\partial \Theta}{\partial t}\nonumber\\
\label{eom}
\end{eqnarray}

and

\begin{eqnarray}
E_{\mu} & = & \frac{\rho_m}{\overline{\rho}} \left\{
       \frac{r}{3}\frac{d}{dr}\left[H_T \frac{d}{dr} \lp
       \frac{\varphi}{\delta} \Lambda \rp - \lp \chi_{\mu} +
       \frac{\varphi}{\delta} \chi_{T}+\frac{\varphi}{\delta}\rp \Lambda
       \right] \right. \\
       & - & \left. \frac{2H_T}{r} \frac{\varphi}{\delta} \Lambda 
       \right\} + 
       \frac{(\overline{\varepsilon}+\overline{\varepsilon}^{grav})}{\varepsilon_m}
       \left[H_T \frac{d}{dr} \lp \frac{\varphi}{\delta} \Lambda \rp
       + f_{\varepsilon} \lp \frac{\varphi}{\delta} \varepsilon_T + 
       \varepsilon_{\mu} \rp \right. \nonumber \\
& -& \left. \chi_{\mu} - \frac{\varphi}{\delta}
       (\chi_T+1)\Lambda \right]. \nonumber 
\label{emu}
\end{eqnarray}
In these expressions we have used the following notations:
 $$H_T = \frac{-d r}{d \ln T}$$ is the temperature scale
height, $$K  = \frac{\chi}{\rho C_p} =
\frac{4acT^3}{3\rho^2 \kappa C_p}$$ is the
thermal diffusivity and $\chi$ is the thermal conductivity.
$\overline{\varepsilon}$ and
$\overline{\varepsilon}^{grav}$ are the means over an isobar of the nuclear and gravitational
energy respectively, and $$f_{\varepsilon} \equiv
\overline{\varepsilon}/(\overline{\varepsilon}+\overline{\varepsilon}^{grav})$$
represents the nuclear fraction of energy. 
$\varepsilon_m$  =$ L(r)/M(r)$ and $\rho_m = M(r)/(4\pi r^3/3)$ are the
mean total energy and the mean density inside a sphere of radius $r$.
$$\chi_\mu = \lp \frac{\partial \ln \chi}{\partial \ln \mu} \rp _{P,T} ~~~~~~~~~~,~~~~~~~~~~~\chi_T = \lp \frac{\partial \ln \chi}{\partial \ln T} \rp_{P,\mu}$$
are logarithmic derivatives of the thermal conductivity $\chi$ ,
and $$\varepsilon_\mu = \lp \frac{\partial \ln \varepsilon}{\partial \ln
\mu} \rp_{P,T}~~~~~~~~~~,~~~~~~~~~~~\varepsilon_T = \lp \frac{\partial \ln
\varepsilon}{\partial \ln T} \rp_{P,\mu}$$ are the logarithmic derivatives of
the nuclear energy.
$$\varphi = \lp \frac{\partial \ln \rho}{\partial \ln \mu} \rp_{P,T}
~~~~~~~~~~,~~~~~~~~~~~\delta = - \lp \frac{\partial \ln \rho}{\partial \ln
  T} \rp_{P,\mu}$$  are derived from the equation of state using the
notation by Kippenhahn \& Weigert (\cite{KW94}).

The term $\frac{\overline{\Omega^{2}}}{2 \pi G \overline{\rho}} \Theta$ in
Eq.~\ref{eom} is of order 2 in $\Omega^2$ whereas the other terms are of
order 1, and is thus neglected.

\end{document}